\documentclass{IEEEtran}

\newif\ifreview

\reviewfalse

\usepackage[]{graphicx}

\usepackage{cite}
\usepackage{amsmath,amssymb,amsfonts}
\usepackage{textcomp}
\usepackage[normalem]{ulem}
\usepackage[hidelinks]{hyperref}
\usepackage[caption=false]{subfig} 

\def\BibTeX{{\rm B\kern-.05em{\sc i\kern-.025em b}\kern-.08em
    T\kern-.1667em\lower.7ex\hbox{E}\kern-.125emX}}

\begin{document}

\title{\vspace{-0.3in}A Chebyshev-Based High-Order-Accurate Integral Equation Solver for Maxwell's Equations}

\author{Jin Hu, \IEEEmembership{Student Member, IEEE}, Emmanuel Garza, \IEEEmembership{Member, IEEE}, and Constantine Sideris, \IEEEmembership{Member, IEEE}
\thanks{\copyright 2021 IEEE.  Personal use of this material is permitted.  Permission from IEEE must be obtained for all other uses, in any current or future media, including reprinting/republishing this material for advertising or promotional purposes, creating new collective works, for resale or redistribution to servers or lists, or reuse of any copyrighted component of this work in other works.}
\thanks{The authors gratefully acknowledge support by the National Science Foundation (1849965), the Air Force Office of Scientific Research (FA9550-20-1-0087), and the National Science Foundation under Grant 2030859 to the Computing Research Association for the CIFellows Project.}
\thanks{J. Hu, E. Garza, and C. Sideris are with the Department of Electrical and Computer Engineering, University of Southern California, Los Angeles, CA 90089, USA (e-mails: jinhu@usc.edu, egarzago@usc.edu, csideris@usc.edu).} \vspace{-0.3in}}

\maketitle

\begin{abstract}
This paper introduces a new method for discretizing and solving integral equation formulations of Maxwell's equations which achieves spectral accuracy for smooth surfaces. The approach is based on a hybrid Nystr\"om-collocation method using Chebyshev polynomials to expand the unknown current densities over curvilinear quadrilateral surface patches. As an example, the proposed strategy is applied the to Magnetic Field Integral Equation (MFIE) and the N-M\"uller formulation for scattering from metallic and dielectric objects, respectively. The convergence is studied for several different geometries, including spheres, cubes, and complex NURBS geometries imported from CAD software, and the results are compared against a commercial Method-of-Moments solver using RWG basis functions.
\end{abstract}


\section{Introduction\label{sec:introduction}}
\IEEEPARstart{D}{ue} to the lack of analytical solutions for anything but the simplest problems~\cite{bowman1987electromagnetic}, efficient and accurate numerical methods for solving Maxwell's equations are crucial for a plethora of engineering applications today, including antennas, microwave devices, and nanophotonic structures. A recent resurgence in inverse design approaches~\cite{lalau2013adjoint}, which involve the automated design of novel electromagnetic structures given a set of desired performance metrics and design constraints, require accurate field and gradient information at each iteration, highlighting the need for fast Maxwell solvers. Although finite difference~\cite{taflove2005computational} and finite element methods~\cite{zienkiewicz1977finite} are popular approaches due to their relative ease of implementation, they suffer from several major drawbacks: poor convergence due to finite difference approximations or low-order basis functions, significant numerical dispersion due to relying on local discrete differentiation, and they are often impractical for large problems due to their volumetric nature. On the other hand, boundary equation (BIE) formulations have been shown to be highly effective in situations containing scatterers with small surface area to volume ratios due to only solving for unknowns on surfaces rather than volumes. Recently, BIEs have been successfully applied towards the modeling and optimization of nanophotonic devices in two dimensions, showing significant improvements in speed and accuracy over finite difference based methods~\cite{sideris2019ultrafast}.

The majority of present day implementations of BIE methods rely on discretization of objects via  triangular discretizations. In the pioneering work by Rao, Wilton and Glisson~\cite{rao1982electromagnetic}, the RWG set of basis functions were introduced in order to solve the Electric Field Integral Equation (EFIE) in conjunction with the Method of Moments (MoM) for flat triangular discretizations. Some of the limitations of RWG functions include that they are only first order and cannot accurately approximate complex surface current distributions without very fine meshing, which often leads to poor convergence and conditioning of the discretized system. Several efforts have been made to improve performance, including the use of alternative basis functions for testing or expansion~\cite{andriulli2012loop}, and use of higher order basis functions~\cite{Wandzura1992,JingguoWang1997,Graglia1997,Hellicar2008,Jorgensen2004}. In particular, \cite{Wandzura1992} extends the RWG basis to curvilinear triangular patches, \cite{JingguoWang1997} presents a p-adaptive scheme for high-order edge basis functions that guarantee continuity of the normal component of the surface currents across elements, and~\cite{Graglia1997} introduced vector basis functions for divergence-conforming and curl-conforming mixed-order N\'{e}d\'{e}lec spaces~\cite{Nedelec1980}. Additionally, other MoM approaches that can handle defective meshes have been proposed, including the  high-order grid-robust method from~\cite{GangKang2001} and the mesh-free scheme from~\cite{Ding2014}.

Other high-order approaches based on Galerkin~\cite{Ganesh2004,Ganesh2009,Ganesh2006,Ganesh2008} and Nystr\"{o}m methods have also been proposed---for example, in~\cite{Canino1998} the singularities in the integral operators are handled by local corrections in the discretization of the kernels. In \cite{Bruno2001b, Bruno2001, bruno2009electromagnetic}, an alternative approach was introduced, which achieves high-order accuracy by utilizing a Nystr\"om method and discretizing the integrals on the basis of local coordinate charts together with fixed and floating partitions of unity. While effective, the approach of \cite{bruno2009electromagnetic} relies on overlapping parameterized patches which can both increase the number of unknowns as well as significantly complicate the generation of surface meshes. Recently, \cite{bruno2018chebyshev} demonstrated a new high-order solution strategy for acoustic scattering problems based on non-overlapping parametric curvilinear patches. The method presented in~\cite{bruno2018chebyshev,Garza2020} discretizes the unknowns on each patch on a Chebyshev grid, approximating the unknown surface densities using Chebyshev polynomials. A spectrally accurate Fej\'er quadrature rule is used for evaluating far interactions, and a Cartesian change of variables is used to cancel the singularity of the integrals associated with local and near interactions (similar in nature to the annihilation procedure described in~\cite{Jorgensen2004}), leading to high-order accuracy in the numerical evaluation of both the singular and near-singular integrals.

In this work, we extend the methods presented in \cite{bruno2018chebyshev,Garza2020} from the scalar Helmholtz equation to the numerical solution of the fully-vectorial Maxwell case, demonstrating that the same integration strategies for non-adjacent, singular, and near-singular interactions work well in the electromagnetic case. In order to demonstrate the generality of the approach, we consider scattering from both Perfect Electrical Conductor (PEC) and dielectric objects. We focus on the solution of the MFIE formulation~\cite{maue1949formulation} for metallic objects and the N-M\"uller formulation\cite{muller2013foundations} for dielectric objects due to their superior conditioning properties, although we remark that all of the methods presented in this work can readily be extended to the Electric and Combined Field Integral Equations (EFIE/CFIE) and other integral equation formulations designed for dielectric objects, such as the Poggio-Chang-Miller-Harrington-Wu-Tsai (PCMHWT) formulation~\cite{yla2008analysis}. 

This paper is organized as follows. In Section~\ref{sec:formulation}, we briefly review the MFIE and the N-M\"uller formulations. In Section~\ref{sec:discretization}, we review the proposed high-order-accurate Chebyshev-based Boundary Integral Equation (CBIE) approach \cite{bruno2018chebyshev, Garza2020} and extend it to the vectorial case necessary for discretizing the integral formulations. Finally, numerical results are presented in Section~\ref{sec:results} which evaluate the performance of the CBIE method by comparing the numerical solutions of plane wave scattering from a PEC/dielectric sphere against analytical Mie-series solutions, as well as solving a PEC/dielectric cube for which no closed-form solutions exist. The accuracy is also compared against a commercial RWG-based MoM solver. Finally, we present results for scattering from two complex NURBS parametrized geometries generated by commercial CAD software.

\section{Integral Equation Formulations\label{sec:formulation}}
\subsection{Magnetic Field Integral Equation Formulation for Closed Metallic Scatterers}
\begin{figure}[t]
\centering
\subfloat[][]{
\includegraphics[width=40mm]{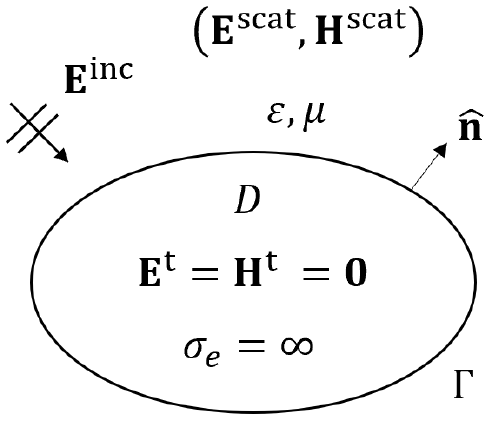}
}
\subfloat[][]{
\includegraphics[width=40mm]{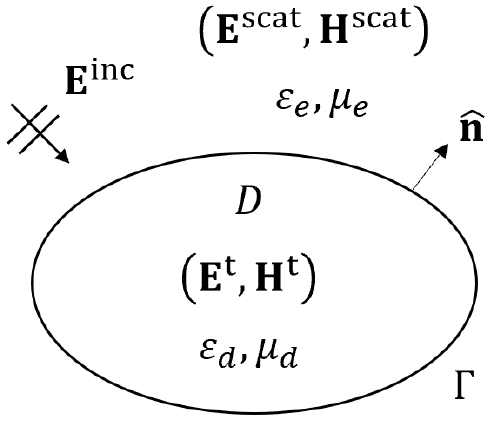}
}
\caption{(a) EM scattering from a closed PEC object. (b) EM scattering from a closed penetrable dielectric object.}
\label{fig:scat_diag}
\end{figure}

We consider the problem of computing the scattered electric and magnetic fields $\left(\mathbf{E}^\text{scat},\mathbf{H}^\text{scat}\right)$ that result due to an incident field excitation $\left(\mathbf{E}^\text{inc},\mathbf{H}^\text{inc}\right)$ impinging on the surface $\Gamma$ of a closed perfect metallic object $D$ as illustrated in  Fig.~\ref{fig:scat_diag}(a). Based on the Stratton-Chu formulas~\cite{volakis2012integral}, Electric and Magnetic Field Integral Equations (EFIE/MFIE) can be derived which express the scattered electric and magnetic fields in terms of the physical current 
$\mathbf{J}=\mathbf{\hat{n}}\times\mathbf{H}$ on the surface of a perfect metallic conducting object~\cite{Nedelec2001}. Although either the EFIE, the MFIE, or a linear combination of the two can be used to solve for the scattered fields due to an incident excitation, only the MFIE is considered in this work due to its good conditioning properties as a result of the nature of Fredholm integral equations of the second kind~\cite{volakis2012integral}. The classical MFIE can be expressed as
\begin{equation}
    \frac{\mathbf{J}}{2} + \mathcal{K}\mathbf{J} = \mathbf{\hat{n}} \times \mathbf{H}^{\text{inc}},
    \label{eq:mfie}
\end{equation}
where $\mathcal{K}$ is the operator:
\begin{equation}
    \mathcal{K}\left[\mathbf{a}\right]\left(\mathbf{r}\right) = \mathbf{\hat{n}}(\mathbf{r}) \times \int_{\Gamma}\mathbf{a}(\mathbf{r'})\times\nabla G(\mathbf{r}-\mathbf{r'})d\sigma(\mathbf{r'}).
\end{equation}
Note that $\nabla$ denotes the gradient with respect to the coordinates of observation points $\mathbf{r}$, $G$ corresponds to the free space scalar Green's function of the Helmholtz equation: $G\left(\mathbf{r}-\mathbf{r}'\right)=\exp\left(-ik\left|\mathbf{r}-\mathbf{r}'\right|\right)/(4\pi\left|\mathbf{r}-\mathbf{r}'\right|)$ with wavenumber $k=2\pi/\lambda$, and $\mathbf{\hat{n}}$ denotes the outwardly pointing surface normal.
\subsection{N-M\"uller Formulation for Dielectric Scatterers}
The second scenario that we consider is scattering from a penetrable dielectric object $D$ with a permittivity $\epsilon_d$ and a permeability $\mu_d$ embedded in a homogeneous background medium characterized by permittivity $\epsilon_e$ and permeability $\mu_e$ in the presence of an incident field excitation $\left(\mathbf{E}^\text{inc},\mathbf{H}^\text{inc}\right)$. As shown in Fig.~\ref{fig:scat_diag}(b), since the object is now penetrable, the incident fields lead to scattered fields outside the object, $\left(\mathbf{E}^\text{scat},\mathbf{H}^\text{scat}\right)$, as well as transmitted fields inside, $\left(\mathbf{E}^\text{t},\mathbf{H}^\text{t}\right)$. Equivalent electric and magnetic current densities can then be defined based on the boundary tangential magnetic and electric fields, respectively, across the dielectric interface as: $\mathbf{J}=\mathbf{\hat{n}}\times(\mathbf{H}^\text{inc}+\mathbf{H}^\text{scat})=\mathbf{\hat{n}}\times\mathbf{H}^\text{t}$ and $\mathbf{M}=(\mathbf{E}^\text{inc}+\mathbf{E}^\text{scat})\times\mathbf{\hat{n}}=\mathbf{E}^\text{t}\times\mathbf{\hat{n}}$ on the surface $\Gamma$ of $D$. By invoking the Stratton-Chu formula for the electric and magnetic fields outside of the object and crossing with the normal vector $\mathbf{\hat{n}}$, we obtain:
\begin{equation}
    \frac{\mathbf{M}}{2} + \mathcal{K}_e\mathbf{M}-\eta_e\mathcal{T}_e\mathbf{J} = -\mathbf{\hat{n}} \times \mathbf{E}^{\text{inc}},
    \label{NEFIE_e}
\end{equation}
\begin{equation}
    \frac{\mathbf{J}}{2} + \mathcal{K}_e\mathbf{J} + \frac{1}{\eta_e}\mathcal{T}_e\mathbf{M} = \mathbf{\hat{n}} \times \mathbf{H}^{\text{inc}},
    \label{NMFIE_e}
\end{equation}
where the $\mathcal{K}_e$ and $\mathcal{T}_e$ operators are defined as:
\begin{equation}
    \mathcal{K}_e\left[\mathbf{a}\right]\left(\mathbf{r}\right) = \mathbf{\hat{n}}(\mathbf{r}) \times \int_{\Gamma}\mathbf{a}(\mathbf{r'})\times\nabla G_e(\mathbf{r}-\mathbf{r'})d\sigma(\mathbf{r'}),
\end{equation}
\begin{equation}
    \mathcal{T}_e\left[\mathbf{a}\right]\left(\mathbf{r}\right) = \mathcal{T}_e^s\left[\mathbf{a}\right]\left(\mathbf{r}\right)+\mathcal{T}_e^h\left[\mathbf{a}\right]\left(\mathbf{r}\right),
\end{equation}
\begin{equation}
    \mathcal{T}_e^s\left[\mathbf{a}\right]\left(\mathbf{r}\right) = jk_e\mathbf{\hat{n}}(\mathbf{r}) \times \int_{\Gamma}\mathbf{a}(\mathbf{r'}) G_e(\mathbf{r}-\mathbf{r'})d\sigma(\mathbf{r'}),
\end{equation}
\begin{equation}
    \mathcal{T}_e^h\left[\mathbf{a}\right]\left(\mathbf{r}\right) =\frac{j}{k_e}\mathbf{\hat{n}}(\mathbf{r}) \times \int_{\Gamma} \nabla G_e(\mathbf{r}-\mathbf{r'})\nabla'_s\cdot\mathbf{a}(\mathbf{r'})d\sigma(\mathbf{r'}),
\end{equation}
where the subscript ``e" in the operators indicates the exterior medium, which has wavenumber $k_e=2\pi/\lambda_e$ and impedance: $\eta_e = \sqrt{\mu_e/\epsilon_e}$.

Similarly, another set of integral equations can be obtained for the transmitted fields $\left(\mathbf{E}^\text{t},\mathbf{H}^\text{t}\right)$ inside the object:
\begin{equation}
    \frac{\mathbf{M}}{2} - \mathcal{K}_d\mathbf{M}+\eta_d\mathcal{T}_d\mathbf{J} = \mathbf{0},
    \label{NEFIE_d}
\end{equation}
\begin{equation}
    \frac{\mathbf{J}}{2} - \mathcal{K}_d\mathbf{J} - \frac{1}{\eta_d}\mathcal{T}_d\mathbf{M} = \mathbf{0},
    \label{NMFIE_d}
\end{equation}
where the $\mathcal{K}_d$ and $\mathcal{T}_d$ operators are defined in the same manner as $\mathcal{K}_e$ and $\mathcal{K}_d$, except the subscript ``d" denotes the interior medium with corresponding wavenumber $k_d = 2\pi/\lambda_d$ and impedance $\eta_d=\sqrt{\mu_d/\epsilon_d}$.

Equations \eqref{NEFIE_e}, \eqref{NMFIE_e}, \eqref{NEFIE_d}, and \eqref{NMFIE_d} give four equations for two unknowns $\left(\mathbf{J},\mathbf{M}\right)$. They can be linearly combined as follows to reduce the system to two independent equations:
\begin{equation}
\begin{aligned}
    \alpha_1\eqref{NEFIE_e}&+\alpha_2\eqref{NEFIE_d}, \\
    \beta_1\eqref{NMFIE_e}&+\beta_2\eqref{NMFIE_d}.
\end{aligned}
\end{equation}
Choosing $\alpha_1 = \epsilon_e,\alpha_2 = \epsilon_d,\beta_1 = \mu_e,\beta_2 = \mu_d$ results in the classical N-M\"uller formulation, which completely cancels the singular terms arising from the gradient of the Green's function in the $\mathcal{T}^h_e$ and $\mathcal{T}^h_d$ operators~\cite{yla2005well}. The combined system in matrix form is thus:
\begin{multline}
    \hspace{-0.1in} \begin{bmatrix}
    \epsilon_e\mathcal{K}_e-\epsilon_d\mathcal{K}_d+\frac{\epsilon_e+\epsilon_d}{2}\mathcal{I} & 
    -(\mathcal{MT}^s+\mathcal{MT}^h)\\
    \mathcal{MT}^s+\mathcal{MT}^h &
    \mu_e\mathcal{K}_e-\mu_d\mathcal{K}_d+\frac{\mu_e+\mu_d}{2}\mathcal{I}
    \end{bmatrix}
    \begin{bmatrix} 
    \mathbf{M} \\
    \mathbf{J} 
    \end{bmatrix} \\
    =\begin{bmatrix} 
    -\epsilon_e\mathbf{\hat{n}} \times \mathbf{E}^{\text{inc}} \\
    \mu_e\mathbf{\hat{n}} \times \mathbf{H}^{\text{inc}}
    \end{bmatrix},
    \label{Nmuller}
\end{multline} 
where $\mathcal{I}$ is the identity operator, and $\mathcal{MT}^s$ and $\mathcal{MT}^h$ are defined as
\begin{multline}
\mathcal{MT}^s \left[\mathbf{a}\right]\left(\mathbf{r}\right) = (\sqrt{\mu_e\epsilon_e}\mathcal{T}_e^s-\sqrt{\mu_d\epsilon_d}\mathcal{T}_d^s)\left[\mathbf{a}\right]\left(\mathbf{r}\right)
\\ = \frac{j}{\omega}\mathbf{\hat{n}}(\mathbf{r}) \times \int_{\Gamma}\mathbf{a}(\mathbf{r'})(k_e^2G_e-k_d^2G_d)d\sigma(\mathbf{r'}),
\end{multline}
\begin{multline}
\mathcal{MT}^h \left[\mathbf{a}\right]\left(\mathbf{r}\right) = (\sqrt{\mu_e\epsilon_e}\mathcal{T}_e^h-\sqrt{\mu_d\epsilon_d}\mathcal{T}_d^h)\left[\mathbf{a}\right]\left(\mathbf{r}\right)
\\ = \frac{j}{\omega}\mathbf{\hat{n}}(\mathbf{r}) \times \int_{\Gamma} (\nabla G_e-\nabla G_d)\nabla'_s\cdot\mathbf{a}(\mathbf{r'})d\sigma(\mathbf{r'}).
\end{multline}
The difference of the hypersingular operators $\mathcal{T}^h$, $\mathcal{MT}^h$ cancels out the highest order singularity, so that $\mathcal{MT}^s + \mathcal{MT}^h$ is only weakly-singular.

\section{Chebyshev-based Boundary Integral Equation Approach\label{sec:discretization}}
\begin{figure}[!t]
\centerline{\includegraphics[width=\columnwidth]{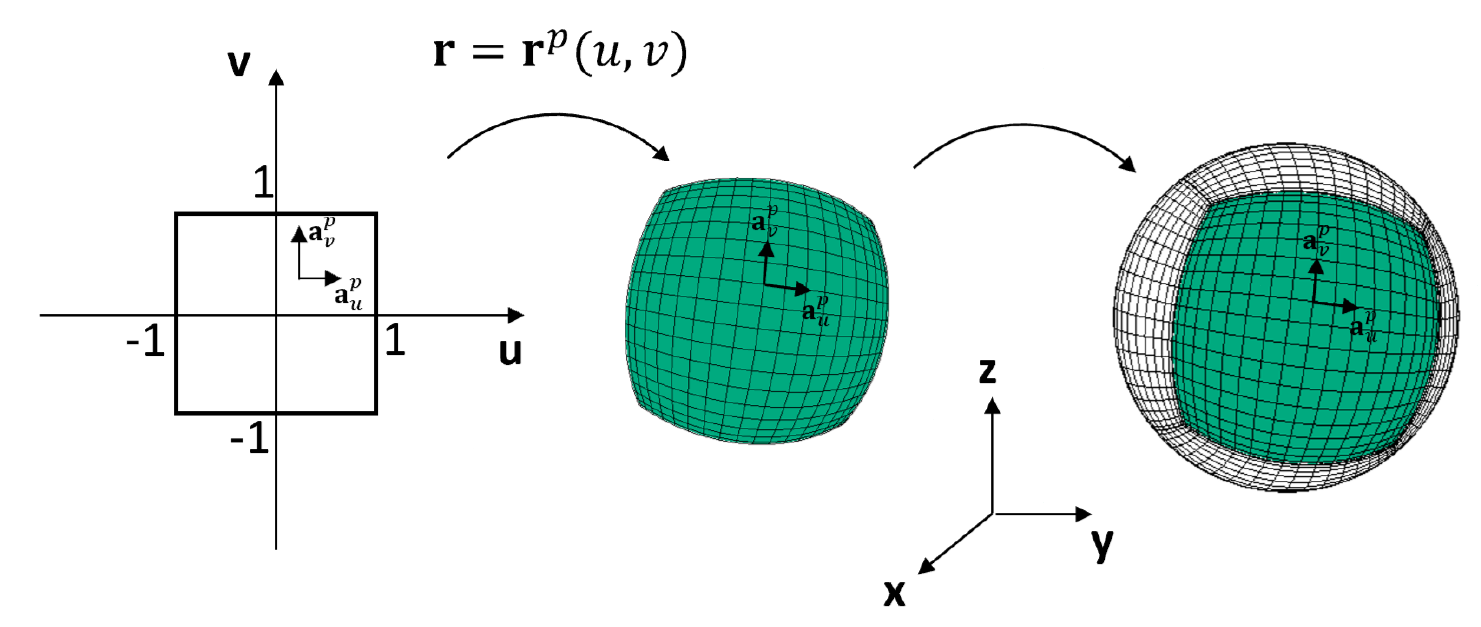}}
\caption{The mapping from square $[-1,1]\times[-1,1]$ in parameter domain to a patch on a sphere in Cartesian coordinates.}
\label{fig:param_mapping}
\end{figure}
\subsection{Representation of Geometries and Densities}
In order to solve \eqref{eq:mfie} or \eqref{Nmuller}, the surface $\Gamma$ is first divided into a number ($M$) of non-overlapping curvilinear quadrilateral patches $\Gamma_p, p=1,2,...,M$. For each of these patches, a $UV$ mapping is used to map from the square $[-1,1]\times[1,1]$ in $UV$ space to the corresponding parameterized surface in Cartesian coordinates as illustrated in Fig.~\ref{fig:param_mapping}. Defining the position vector on $\Gamma_p$ as $\mathbf{r} = \mathbf{r}^p(u,v) = \left(x^p(u,v),y^p(u,v),z^p(u,v)\right)$, we can define the tangential covariant basis vectors and surface normal on $\Gamma_p$ as
\begin{equation}
    \mathbf{a}^p_u = \frac{\partial\mathbf{r}^p(u,v)}{\partial u},
    \; 
    \mathbf{a}^p_v = \frac{\partial\mathbf{r}^p(u,v)}{\partial v},
    \; 
    \mathbf{\hat{n}}^p = \frac{\mathbf{a}^p_u \times \mathbf{a}^p_v}{||\mathbf{a}^p_u \times \mathbf{a}^p_v||}.
\end{equation}
Thus, the vector triplet $\left(\mathbf{a}^p_u,\mathbf{a}^p_v,\mathbf{\hat{n}}^p\right)$ forms a local conformal reference frame at each point on $\Gamma_p$. The metric tensor is defined as
\begin{equation}
    G^p = \begin{bmatrix}
    g^p_{uu} & g^p_{uv}\\
    g^p_{vu} & g^p_{vv}\end{bmatrix},
\end{equation}
where $g^p_{ij} = \mathbf{a}^p_{i}\cdot\mathbf{a}^p_{j}$ and thus we have a surface element Jacobian $ds=\sqrt{|G^p|}\text{d}u \text{d}v$ on $\Gamma_p$ where $|G^p|$ is the determinant of $G^p$. We can now represent the surface current densities on $\Gamma_p$ as
\begin{align}
   \mathbf{J}^p (u,v)
    &= \frac{J^{p,u}(u,v)\mathbf{a}^p_u(u,v) + J^{p,v}(u,v)\mathbf{a}^p_v(u,v)}
    {\sqrt{|G^p(u,v)|}},
    \label{Jpdef} \\
     \hspace{-0.1in}\mathbf{M}^p (u,v) 
    &= \frac{M^{p,u}(u,v)\mathbf{a}^p_u(u,v) + M^{p,v}(u,v)\mathbf{a}^p_v(u,v)}
    {\sqrt{|G^p(u,v)|}},
\end{align}
for $p=1,...,M$, where $\mathbf{J}^p (u,v) \equiv \mathbf{J} (\mathbf{r}^p(u,v)) $, $\mathbf{M}^p (u,v) \equiv \mathbf{M} (\mathbf{r}^p(u,v)) $, $J^{p,u}$ (resp. $M^{p,u}$) and $J^{p,v}$ (resp. $M^{p,v}$) are scalar functions representing the contravariant components of the surface current density $\mathbf{J}$ (resp. $\mathbf{M}$) on the $p^{\text{th}}$ patch normalized by the metric tensor, $\sqrt{\left|G^p\right|}$. The densities are normalized by the surface element Jacobian in order to simplify the numerical computation of their divergence (see \cite[sec. 6.2.5]{volakis2012integral}). Due to their desirable spectral convergence properties for approximating smooth functions, we utilize Chebyshev polynomials to discretize the surface current densities:
\begin{equation}
    J^{p,a} = \sum_{m=0}^{N^p_v-1} \sum_{n=0}^{N^p_u-1} \gamma^{p,a}_{n,m}T_n(u)T_m(v),\quad \text{for $a=u,v$}
\label{eq:Jdiscr}
\end{equation}
\begin{equation}
    M^{p,a} = \sum_{m=0}^{N^p_v-1} \sum_{n=0}^{N^p_u-1} \zeta^{p,a}_{n,m}T_n(u)T_m(v),\quad \text{for $a=u,v$}
\label{eq:Mdiscr}
\end{equation}
where the Chebyshev coefficients $\gamma^{p,a}_{n,m}$ and $\zeta^{p,a}_{n,m}$ can be computed from the values of the densities on Chebyshev nodes,
\begin{equation}
    \gamma^{p,a}_{n,m} = \frac{\alpha_n\alpha_m}{N^p_uN^p_v}\sum_{k=0}^{N^p_v-1} \sum_{l=0}^{N^p_u-1} J^{p,a}(u_l,v_k)T_n(u_l)T_m(v_k),
\end{equation}
\begin{equation}
    \zeta^{p,a}_{n,m} = \frac{\alpha_n\alpha_m}{N^p_uN^p_v}\sum_{k=0}^{N^p_v-1} \sum_{l=0}^{N^p_u-1} M^{p,a}(u_l,v_k)T_n(u_l)T_m(v_k),
\end{equation}
based on the discrete orthogonality property of Chebyshev polynomials~\cite{mason2002chebyshev}, with $\alpha_n = 1$ for $n=0$ and $\alpha_n = 2$ otherwise.
Therefore, only the unknowns at the Chebyshev nodes (\ref{eq:chebynodes}) are required to represent the continuous scalar densities $J^{p,a}$ and $M^{p,a}$ over the whole patch $\Gamma_p$, where $a$ can be either $u$ or $v$. 

In our specific implementation, these unknowns are ordered in vector form as:
\begin{multline}
    \mathcal{J}^p = [ 
    J^{p,u}(u_0,v_0),
    \dots ,
    J^{p,u}(u_{N^p_u-1},v_{N^p_v-1}), \\ 
    J^{p,v}(u_0,v_0), 
    \dots ,
    J^{p,v}(u_{N^p_u-1},v_{N^p_v-1})]^T
    \label{eq:Jp}
\end{multline}
and a similar expression holds for $\mathcal{M}^p$.

\subsection{Discretization of Operators}
We now turn our attention towards discretization of the $\mathcal{K}/\mathcal{K}_e/\mathcal{K}_d$, $\mathcal{MT}^s$ and $\mathcal{MT}^h$ operators. We will begin by discretizing the $\mathcal{K}$ operator first. Clearly, any integral over $\Gamma$ can be split into the sum of integrals over each of the $M$ patches, 
\begin{gather}
    \mathcal{K}\left[\mathbf{J}\right](\mathbf{r}) =
    \sum_{p=1}^M \mathcal{K}[\mathbf{J}^p] (\mathbf{r}), \\
\begin{split} \label{eq:Ksplit}
    \mathcal{K}\left[\mathbf{J}^p\right](\mathbf{r}) =
    \mathbf{\hat{n}}(\mathbf{r}) \times \int_{\Gamma^p}\mathbf{J}^p(\mathbf{r'})\times\nabla G(\mathbf{r}-\mathbf{r'}) d\sigma(\mathbf{r'})\\
    =  \mathbf{\hat{n}}(\mathbf{r}) \times \int_{-1}^{1}\int_{-1}^{1} (J^{p,u}(u,v)\mathbf{a}^p_u(u,v) + \\
    J^{p,v}(u,v)\mathbf{a}^p_v(u,v)) \times\nabla G(\mathbf{r}-\mathbf{r}^p(u,v)) \text{d}u \text{d}v.
\end{split}
\end{gather}
Note that the $\sqrt{|G^p(u,v)|}$ in the denominator of the expansion \eqref{Jpdef} for $\mathbf{J}$ cancels with the Jacobian $\sqrt{|G^p(u,v)|}$ that appears in the integral.
In its current form, \eqref{eq:Ksplit} contains the hypersingular kernel $\nabla G$; however, it can be manipulated using the BAC-CAB vector identity into
\begin{multline}
    \hspace{-0.1in} \mathcal{K}\left[\mathbf{J}^p\right](\mathbf{r}) = \int_{-1}^{1}\int_{-1}^{1} J^{p,u}(u,v) \bigg ( \mathbf{a}^p_u(u,v) \frac{\partial G(\mathbf{r}-\mathbf{r}^p(u,v))}{\partial \mathbf{\hat{n}}(\mathbf{r})}\\
    -\nabla G(\mathbf{r}-\mathbf{r}^p(u,v)) \mathbf{\hat{n}}(\mathbf{r})\cdot\mathbf{a}^p_u(u,v) \bigg) \text{d}u \text{d}v + \\
    \int_{-1}^{1}\int_{-1}^{1} J^{p,v}(u,v) \bigg( \mathbf{a}^p_v(u,v) \frac{\partial G(\mathbf{r}-\mathbf{r}^p(u,v))}{\partial \mathbf{\hat{n}}(\mathbf{r})}\\
    -\nabla G(\mathbf{r}-\mathbf{r}^p(u,v)) \mathbf{\hat{n}}(\mathbf{r})\cdot\mathbf{a}^p_v(u,v) \bigg) \text{d}u \text{d}v,
\label{eq:Ksmooth}
\end{multline}
which is weakly singular since $\mathbf{\hat{n}}(\mathbf{r})\cdot\mathbf{a}^p_{u,v}$ approaches 0 as $\mathbf{r}^p(u,v)\to\mathbf{r}$.  Substituting $\eqref{eq:Ksmooth}$ into $\eqref{eq:mfie}$, we must obtain $2\sum_{p=1}^M N^p_u N^p_v$ linearly independent equations in order to obtain a uniquely solvable linear system for approximating $\mathbf{J}$ on $\Gamma$. This is achieved by using a collocation method and testing \eqref{eq:mfie} at same points as the unknowns.

To obtain the contravariant components of the vector equations~\eqref{eq:mfie} and~\eqref{Nmuller}, we dot each vector equation with the normalized contravariant basis vectors $\sqrt{G^p}\mathbf{a}^{p,u}$ and $\sqrt{G^p}\mathbf{a}^{p,v}$ where the contravariant basis vectors $\mathbf{a}^{p,u}$ and $\mathbf{a}^{p,v}$ are defined via the orthogonality relation
\begin{equation}
    \mathbf{a}^{p,a} \cdot \mathbf{a}^p_b = \begin{cases}
    1 & a=b\\
    0 & a\neq b
    \end{cases}.
  \end{equation}

We can now define the linear system:
\begin{equation}
    \begin{bmatrix}
    \frac{I}{2} + K^{11} & \dots & K^{1M} \\
    \vdots & \vdots & \ddots & \vdots \\
    K^{M1} & \dots & \frac{I}{2} + K^{MM} 
    \end{bmatrix}
    \begin{bmatrix}
    \mathcal{J}^1 \\
    \vdots \\
    \mathcal{J}^M
    \end{bmatrix} =
    \begin{bmatrix}
    \mathcal{H}^{1}_{\text{inc}} \\
    \vdots \\
    \mathcal{H}^{M}_{\text{inc}}
    \end{bmatrix},
    \label{MFIE_mat}
\end{equation}
where
\begin{equation}
\begin{aligned} 
\mathcal{H}_{\text{inc}}^p = [ &
-\mathbf{a}^p_v\cdot\mathbf{H}^{p,\text{inc}}(u_0,v_0) ,
\dots , \\ &
-\mathbf{a}^p_v\cdot\mathbf{H}^{p,\text{inc}}(u_{N^p_u-1},v_{N^p_v-1}) , \\
& \mathbf{a}^p_u\cdot\mathbf{H}^{p,\text{inc}}(u_0,v_0) ,
\dots , \\ &
\mathbf{a}^p_u\cdot\mathbf{H}^{p,\text{inc}}(u_{N^p_u-1},v_{N^p_v-1}) ]^T
\end{aligned}
\label{rhs}
\end{equation}
represents the incident magnetic field on the $p^{\text{th}}$ patch and $\mathcal{J}^p, p=1,2,..,M$ is given by~\eqref{eq:Jp}.

The matrix block $K^{qp}$ represents contributions of the appropriately discretized $\mathcal{K}$ operator from the densities of the patch $p$ to the target points on patch $q$ and consists of the individual sub-blocks:
\begin{equation}
    K^{qp} = \begin{pmatrix}
    K^{qp}_{uu} & K^{qp}_{uv} \\
    K^{qp}_{vu} & K^{qp}_{vv}
    \end{pmatrix}.
    \label{Kblock}
\end{equation}

For the operators used in the N-M\"uller formulation, the matrix blocks corresponding to the $\mathcal{K}_e$ and $\mathcal{K}_d$ operator can be obtained in exactly the same way as those for the $\mathcal{K}$ operator by simply replacing the wavenumber $k$ in the Green's function in \eqref{eq:Ksmooth} with $k_e$ and $k_d$ respectively. The integral of the $\mathcal{MT}^s$ and $\mathcal{MT}^h$ operators can also be split over each patch in a similar way as the $\mathcal{K}$ operator: 
\begin{gather}
    \mathcal{MT}^s\left[\mathbf{J}\right](\mathbf{r}) = \sum_{p=1}^M \mathcal{MT}^s[\mathbf{J}^p](\mathbf{r}), \\
\begin{split}
    \mathcal{MT}^s\left[\mathbf{J}^p\right](\mathbf{r}) = \frac{j}{\omega}\mathbf{\hat{n}}(\mathbf{r}) \times \int_{\Gamma^p}\mathbf{J}^p(\mathbf{r'})G^p_\Delta d\sigma(\mathbf{r'})\\
    = \frac{j}{\omega}\mathbf{\hat{n}}(\mathbf{r}) \times \int_{-1}^{1}\int_{-1}^{1} \big( J^{p,u}(u,v)\mathbf{a}^p_u(u,v) + \\
    J^{p,v}(u,v)\mathbf{a}^p_v(u,v) \big) G^p_\Delta(\mathbf{r},u,v)   \text{d}u \text{d}v,
\label{eq:MTssplit}
\end{split}
\end{gather}
\begin{gather}
    \mathcal{MT}^h \left[\mathbf{J}\right](\mathbf{r}) = \sum_{p=1}^M \mathcal{MT}^h[\mathbf{J}^p](\mathbf{r}),\\
\begin{split}
    \mathcal{MT}^h \left[\mathbf{J}^p\right](\mathbf{r}) = \frac{j}{\omega}\mathbf{\hat{n}}(\mathbf{r}) \times \int_{\Gamma^p} 
    \nabla G^p_\Delta \nabla'_s\cdot\mathbf{J}^p(\mathbf{r'})d\sigma(\mathbf{r'})\\
    = \frac{j}{\omega}\mathbf{\hat{n}}(\mathbf{r}) \times \int_{-1}^{1}\int_{-1}^{1} \nabla G^p_\Delta \bigg( \frac{\partial J^{p,u}}{\partial u}+\frac{\partial J^{p,v}}{\partial v} \bigg) \text{d}u \text{d}v \\
    = \frac{j}{\omega}\mathbf{\hat{n}}(\mathbf{r}) \times \int_{-1}^{1}\int_{-1}^{1} \nabla G^p_\Delta(\mathbf{r},u,v) \sum_{m=0}^{N^p_v-1} \sum_{n=0}^{N^p_u-1} \\
    \big( \gamma^{p,u}_{n,m}T_n'(u)T_m(v)+  \gamma^{p,v}_{n,m}T_n(u)T_m'(v) \big) \text{d}u \text{d}v \\
    = \sum_{m=0}^{N^p_v-1} \sum_{n=0}^{N^p_u-1} \frac{j}{\omega}\mathbf{\hat{n}}(\mathbf{r}) \times \int_{-1}^{1}\int_{-1}^{1} \big( \gamma^{p,u}_{n,m}T_n'(u)T_m(v)+\\
    \gamma^{p,v}_{n,m}T_n(u)T_m'(v) \big) \nabla G^p_\Delta(\mathbf{r}, u, v) \text{d}u \text{d}v, 
\label{eq:MThsplit}
\end{split}
\end{gather}
where $ G^p_\Delta(\mathbf{r},u,v) \equiv [k_e^2G_e(\mathbf{r}-\mathbf{r}^p(u,v))-k_d^2G_d(\mathbf{r}-\mathbf{r}^p(u,v))]$ and $\nabla G^p_\Delta(\mathbf{r},u,v) \equiv [\nabla G_e(\mathbf{r}-\mathbf{r}^p(u,v))-\nabla G_d(\mathbf{r}-\mathbf{r}^p(u,v))]$.
The partial derivative of the densities can be readily computed by taking the derivative of the corresponding Chebyshev polynomials~\cite{Press2007}. After the substitution of \eqref{eq:MTssplit} and \eqref{eq:MThsplit} into \eqref{Nmuller} with the expansion defined in \eqref{eq:Jdiscr} and \eqref{eq:Mdiscr}, testing \eqref{Nmuller} at the same collocation points as the unknowns results in the linear system:
\begin{multline}
    \hspace{-0.1in}  \begin{bmatrix}
    \epsilon_e K_e-\epsilon_d K_d+\frac{\epsilon_e+\epsilon_d}{2}I &
    -(MT^s+MT^h)\\
    MT^s+MT^h &
    \mu_e K_e-\mu_d K_d+\frac{\mu_e+\mu_d}{2}I
    \end{bmatrix}
    \begin{bmatrix} 
    \mathcal{M} \\
    \mathcal{J} 
    \end{bmatrix} \\
    =\begin{bmatrix} 
    -\epsilon_e\mathcal{E}_{\text{inc}}  \\
    \mu_e\mathcal{H}_{\text{inc}}
    \end{bmatrix} .
\label{Nmuller_mat}
\end{multline}
The block in $\mathcal{E}_{\text{inc}}$ corresponding to the incident electric field on the $p^{\text{th}}$ patch is: 
\begin{equation}
\begin{aligned} 
\mathcal{E}_{\text{inc}}^p = [&
-\mathbf{a}^p_v\cdot\mathbf{E}^{p,\text{inc}}(u_0,v_0) ,
\dots , \\ &
-\mathbf{a}^p_v\cdot\mathbf{E}^{p,\text{inc}}(u_{N^p_u-1},v_{N^p_v-1}) ,\\ &
\mathbf{a}^p_u\cdot\mathbf{E}^{p,\text{inc}}(u_0,v_0) ,
\dots , \\ &
\mathbf{a}^p_u\cdot\mathbf{E}^{p,\text{inc}}(u_{N^p_u-1},v_{N^p_v-1}) ]^T.
\end{aligned} 
\label{erhs}
\end{equation}
The counterpart $\mathcal{H}_{\text{inc}}^p$ is defined in \eqref{rhs}. The matrices $K_e$, $K_d$, $MT^s$ and $MT^h$ all have the same block structure arranged by patches as indicated in \eqref{MFIE_mat} and \eqref{Kblock} for the matrix $K$.
A suitable numerical integration strategy must now be chosen for evaluating the necessary operators to compute the above matrix sub-blocks. In the following two subsections, we will detail the approach for dealing with the non-adjacent interactions ($p \neq q$) and the singular and near-singular interactions arising either when $p=q$ or when $p \neq q$, but the target point on $q$ is located very near to the source patch $p$, which is based on the strategy put forth in~\cite{bruno2018chebyshev}.

\subsection{Non-Adjacent Interactions}
The integrals \eqref{eq:Ksmooth}, \eqref{eq:MTssplit} and \eqref{eq:MThsplit} are smooth for target points far away from the source patch $p$. Since the current density $\mathbf{J}/\mathbf{M}$ is discretized on a Chebyshev grid on each patch, we can use Fej\'er's first quadrature rule to numerically evaluate these integrals with high-order accuracy. The quadrature nodes and weights for an order $N$ open rule are given by:
\begin{equation}
x_i = \cos\left(\pi\frac{2i+1}{2N}\right),\quad i=0,...,N-1, 
\label{eq:chebynodes}
\end{equation}
\begin{equation}
w_i = \frac{2}{N}\left(1-2\sum_{k=1}^{N/2}\frac{1}{4k^2-1}\cos\left(k\pi\frac{2i+1}{N}\right)\right), 
\end{equation}
and the discretized versions of \eqref{eq:Ksmooth}, \eqref{eq:MTssplit} and \eqref{eq:MThsplit} become (with $a=\{u,v\}$ and $b=\{u,v\}$ to represent the $u$ and $v$ contravariant components):
    \begin{multline}
    K^{qp}_{ba}\left[J^{p,a}\right](u',v') =\sum_{k=0}^{N^p_v-1}\sum_{l=0}^{N^p_u-1}A^{qp}_{ba}(u',v',u_l,v_k)\\
    \sqrt{|G^q(u',v')|}w_l w_k J^{p,a}(u_l,v_k) ,
    \end{multline}
\begin{multline}
    MT^{s,qp}_{ba}\left[J^{p,a}\right](u',v') =\sum_{k=0}^{N^p_v-1}\sum_{l=0}^{N^p_u-1}B^{qp}_{ba}(u',v',u_l,v_k)\\
    \sqrt{|G^q(u',v')|}w_l w_k J^{p,a}(u_l,v_k) ,
    \end{multline}
\begin{multline}
    MT^{h,qp}_{ba}\left[J^{p,a}\right](u',v') =\sum_{k=0}^{N^p_v-1}\sum_{l=0}^{N^p_u-1}C^{qp}_{ba}(u',v',u_l,v_k)\\
    \sqrt{|G^q(u',v')|}w_l w_k \frac{\partial J^{p,a}}{\partial a}(u_l,v_k) ,
    \end{multline}
with
\begin{multline}
A^{qp}_{ba}(u',v',u_l,v_k)=\mathbf{a}^{q,b}(u',v')\cdot\mathbf{a}^p_a(u_l,v_k) \\
\frac{\partial G\left(\mathbf{r}^q(u',v')-\mathbf{r}^p(u_l,v_k)\right)}{\partial\mathbf{\hat{n}}^q(u',v')}-\mathbf{\hat{n}}^q(u',v')\cdot\mathbf{a}^p_a(u_l,v_k)\\
\mathbf{a}^{q,b}(u',v')\cdot\nabla G\left(\mathbf{r}^q(u',v')-\mathbf{r}^p(u_l,v_k)\right) ,
\end{multline}
\begin{multline}
B^{qp}_{ba}(u',v',u_l,v_k)=\frac{j}{\omega}\mathbf{a}^{q,b}(u',v')\cdot(\mathbf{\hat{n}}^q(u',v')\times\mathbf{a}^p_a(u_l,v_k))\\
\left[k_e^2G_e-k_d^2G_d\right]\left(\mathbf{r}^q(u',v')-\mathbf{r}^p(u_l,v_k)\right) ,
\end{multline}
\begin{multline}
C^{qp}_{ba}(u',v',u_l,v_k)=\frac{j}{\omega}\mathbf{a}^{q,b}(u',v')\cdot\mathbf{\hat{n}}^q(u',v')\times\\
\left[\nabla G_e -\nabla G_d\right]\left(\mathbf{r}^q(u',v')-\mathbf{r}^p(u_l,v_k)\right) ,
\end{multline}
where $u_l$ and $v_k$ are the discretization points on the Chebyshev grid corresponding to the $x_i$ nodes: $\left.u_l = x_l\right|l=0,\dots,N^p_u-1,\left.v_k = x_k\right|k=0,\dots,N^p_v-1$, and $w_l$ and $w_k$ are the quadrature weights in the $u$ and $v$ directions respectively.

\subsection{Singular and Near-Singular Interactions}
When the observation point $(u',v')$ is on the same patch as the source patch $p$, the integrals \eqref{eq:Ksmooth}, \eqref{eq:MTssplit} and \eqref{eq:MThsplit} become singular\footnote{Actually, the integral $\eqref{eq:MThsplit}$ for $MT^h$ remains regular due to the M\"uller cancellation and does not require special consideration; however, for simplicity we treat it in the same way as the other operators in our implementation.}. In order to accurately compute the resulting integrals with high-order accuracy we consider the following smoothing change of variables~\cite[Sec. 3.5]{colton1998inverse},~\cite{bruno2018chebyshev}
\begin{equation}
\begin{aligned}
    &u(s) = \xi_{u'}(s),\quad &v(t) = \xi_{v'}(t),\quad \text{for $-1\leq s,t \leq 1$}, 
\end{aligned}
\end{equation}
where 
\begin{equation}
\begin{aligned}
    &\xi_\alpha(\tau)=\begin{cases}
    \alpha+\left(\frac{\text{sgn}(\tau)-\alpha}{\pi}\right)w(\pi|\tau|),\quad \text{for } \alpha\neq\pm 1 \\
    \alpha\mp\left(\frac{1\pm a}{\pi}\right)w\left(\pi\left|\frac{\tau\mp 1}{2}\right|\right),\quad \text{for } \alpha = \pm 1
    \end{cases} , \\
    &w(\tau) = 2\pi\frac{\left[\nu(\tau)\right]^d}{\left[\nu(\tau)\right]^d+\left[\nu\left(2\pi-\tau\right)\right]^d},\quad 0\leq\tau\leq2\pi  , \\
    &\nu(\tau) = \left(\frac{1}{d}-\frac{1}{2}\right)\left(\frac{\pi-\tau}{\pi}\right)^3+\frac{1}{d}\left(\frac{\tau-\pi}{\pi}\right)+\frac{1}{2} .
\end{aligned}
\label{change_variable}
\end{equation}
The derivatives of $w(\tau)$ vanish up to order $d-1$ at the endpoints, and therefore $d-1$ derivatives of $\xi_\alpha(\tau)$ also vanish at $\tau=0$, corresponding to $\xi_\alpha(0)=\alpha$. Now, since $J^{p,a}(a=u,v)$ is expanded in terms of Chebyshev polynomials, which satisfy a discrete orthogonality property on the Chebyshev grid points, we can accurately precompute the action of the $\mathcal{K}^{qp}_{ba}$, $\mathcal{MT}^{s,qp}_{ba}$ and $\mathcal{MT}^{h,qp}_{ba}$ operators on each Chebyshev polynomial individually:
\begin{multline}
    K^{qp}_{ba}\left[T_{mn}\right](u',v') =
    \sqrt{|G^q(u',v')|} 
    \sum_{k=0}^{N_\beta^v-1}\sum_{l=0}^{N_\beta^u-1} w_l w_k
    \\ 
    A^{qp}_{ba}(u',v',\xi_{u'}(s_l),\xi_{v'}(t_k))\\
    \frac{\partial u}{\partial s}(s_l)\frac{\partial v}{\partial t}(t_k) T_{mn}(\xi_{u'}(s_l),\xi_{v'}(t_k)),
    \label{eq:precompT}
\end{multline}
where $T_{mn}(u,v) \equiv T_n(u) T_m(v)$, and where $\frac{\partial u}{\partial s}\to 0$ and $\frac{\partial v}{\partial t}\to 0$ as $\xi_{u'}(s)\to u'$ and $\xi_{v'}(t)\to v'$ respectively, canceling the singularity in $A$ up to a degree $d-1$. Note that the expressions for $\mathcal{MT}^{s,qp}_{ba}$ and $\mathcal{MT}^{h,qp}_{ba}$ are the same but with $A$ replaced by $B$ and $C$ respectively. It is important that $N^{u,v}_\beta$ is chosen sufficiently large to accurately compute each of the precomputation integrals in \eqref{eq:precompT} above. A numerical analysis of the resulting forward map accuracy vs. $N^{u,v}_\beta$ is done in Section~\ref{sec:results}. Finally, on the basis of these precomputations, the action of each of these operators on any $J^{p,a}$ or $M^{p,a}$ can be readily computed using the Chebyshev expansion of the density, e.g.
\begin{equation}
    \hspace{-0.1in} K^{qp}_{ba}\left[J^{p,a}\right](u',v') = \sum_{m=0}^{N^p_v-1}\sum_{n=0}^{N^p_u-1} \gamma^{p,a}_{m,n} K^{qp}_{ba}\left[T_{mn}\right](u',v') ,
\end{equation}
where $\gamma^{p,a}_{m,n}$ are the Chebyshev expansion coefficients defined in \eqref{eq:Jdiscr}. An analogous relation also holds true for the $\mathcal{MT}^s$ and $\mathcal{MT}^h$ operators. This precomputation approach is also used in order to accurately compute the $K^{qp}_{ba}$, $MT^{s,qp}_{ba}$ and $MT^{h,qp}_{ba}$ blocks corresponding to target points which are on different patches but which are still in close proximity to the source patch, making the integration near-singular. The only difference in this scenario arises in the selection of $\alpha$ in the change of variable expression \eqref{change_variable}. Instead of simply choosing the $(u',v')$ corresponding to the target point, since it is on a different patch, we search for: 
\begin{equation}
(u^{*},v^{*}) =\mathop{\arg\min} _{(u,v)\in[-1,1]^2} |\mathbf{r}^q(u',v')-\mathbf{r}^p(u,v)| ,
\end{equation}
 for the change-of-variables as the point on the source patch nearest to the target patch, which can be readily found by an appropriate minimization algorithm. We adopted the golden section search algorithm in our specific implementation~\cite{kiefer1953sequential,Press2007}, with initial bounds given by the points on the grid of the source patch which minimize the distance, then using the golden section search to improve that initial guess. We found, just as in~\cite{bruno2018chebyshev}, that this approach is robust and does not incur significant computational expense since it is only performed while precomputing the action of the operators onto the Chebyshev polynomials from~\eqref{eq:precompT}.

As in~\cite{bruno2018chebyshev}, the computational cost of the singular and near-singular integrals (for $N^p_u=N^p_v=N$ and $N_\beta^u=N_\beta^v=N_\beta$) is given by $\mathcal{O}(M N_\beta^2 N (N^2 + N_{\text{close}}))$, where $N_{\text{close}}$ represents the number of points per patch that require near-singular integrations. This bound is obtained by performing the precomputations on~\eqref{eq:precompT} via partial summation~\cite[Sec.~10.2]{Boyd2001}, and it differs only from the acoustic case by a constant factor given that multiple integrals of kernels against the Chebyshev polynomials need to be precomputed, while the acoustic case only involves one kernel. For an implementation that relies on an iterative linear algebra solver, where the matrices are not explicitly formed, the storage of the precomputations require $\mathcal{O}(M N^2 (N^2 + N_{\text{close}}))$ complex-valued numbers. Hence, in practice one must consider a balance between $M$, $N$ and $N_\beta$: a large value of $N$ will give a higher order expansion of the currents, but will incur in larger storage and precomputation times. On the other hand, increasing the number of patches $M$ while keeping $N$ constant results in only linear growth in the storage and precomputation times needed, at the cost of lower polynomial representations of the current densities.

\section{Numerical Results\label{sec:results}}
We first present the convergence of the forward map---namely, the action of the discretized integral operators on a given set of currents---for both the MFIE and N-M\"uller formulations with respect to the number of points per patch per dimension $N$ ($N_u = N_v = N$, corresponding to polynomial representations of the current densities of order $N-1$, as can be seen from~\eqref{eq:Jdiscr} and~\eqref{eq:Mdiscr}) for varying levels of singular integration refinement $N_{\beta}$. Following this, several numerical examples involving scattering from PEC and dielectric spheres and cubes are presented and compared against a commercial RWG-based MoM solver to demonstrate the high accuracy that can be achieved using the proposed CBIE method. Finally, we present scattering and near-field density results from scattering by highly intricate 3D NURBS objects parametrized with commercial CAD software~\cite{rhino_cite}, which shows that the approach can be readily applied to simulate objects arising in realistic applications.

\subsection{Forward Map Convergence}
\begin{figure}[ht]
\centering
\subfloat[][]{
\includegraphics[width=80mm]{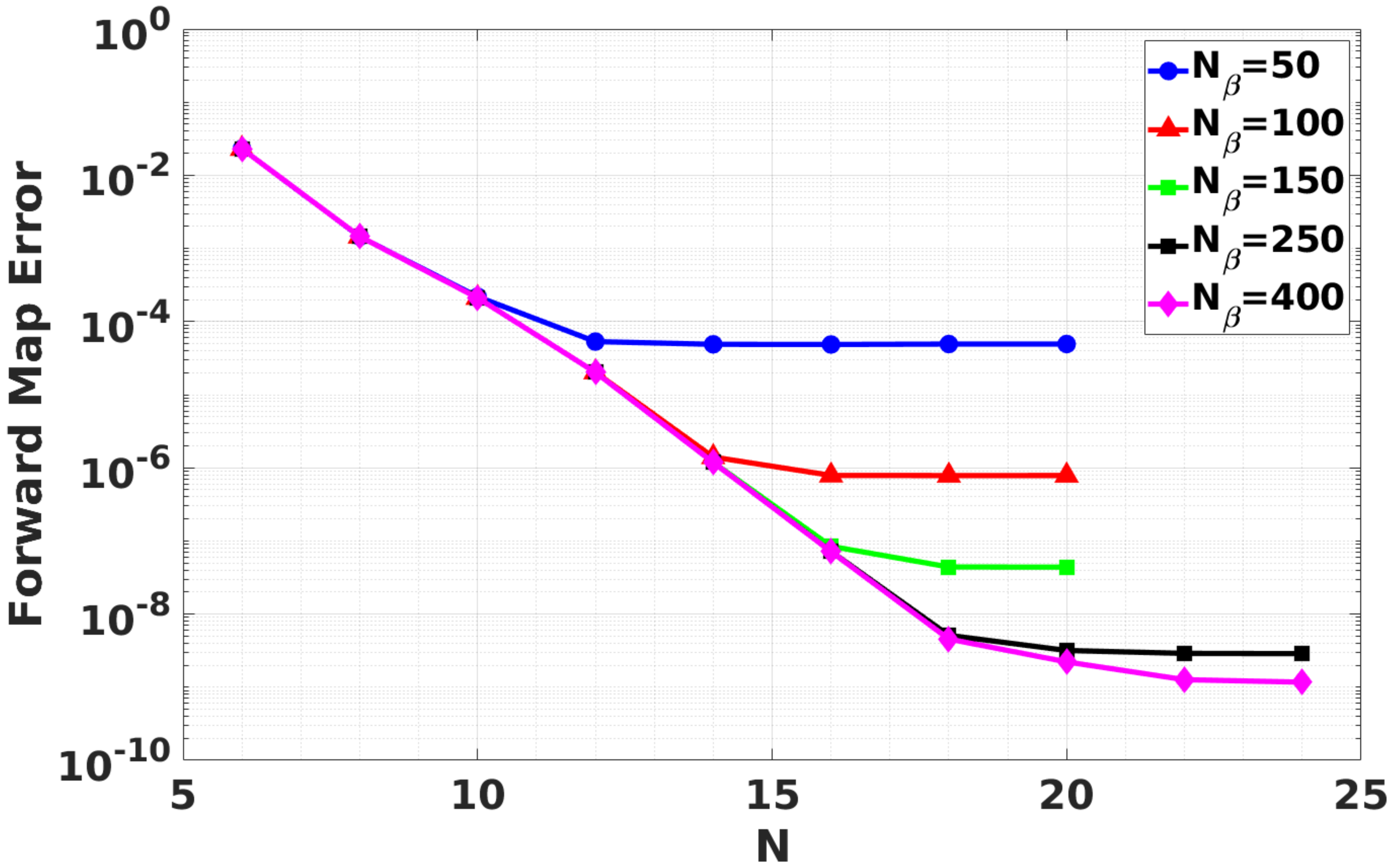}
}\\
\subfloat[][]{
\includegraphics[width=80mm]{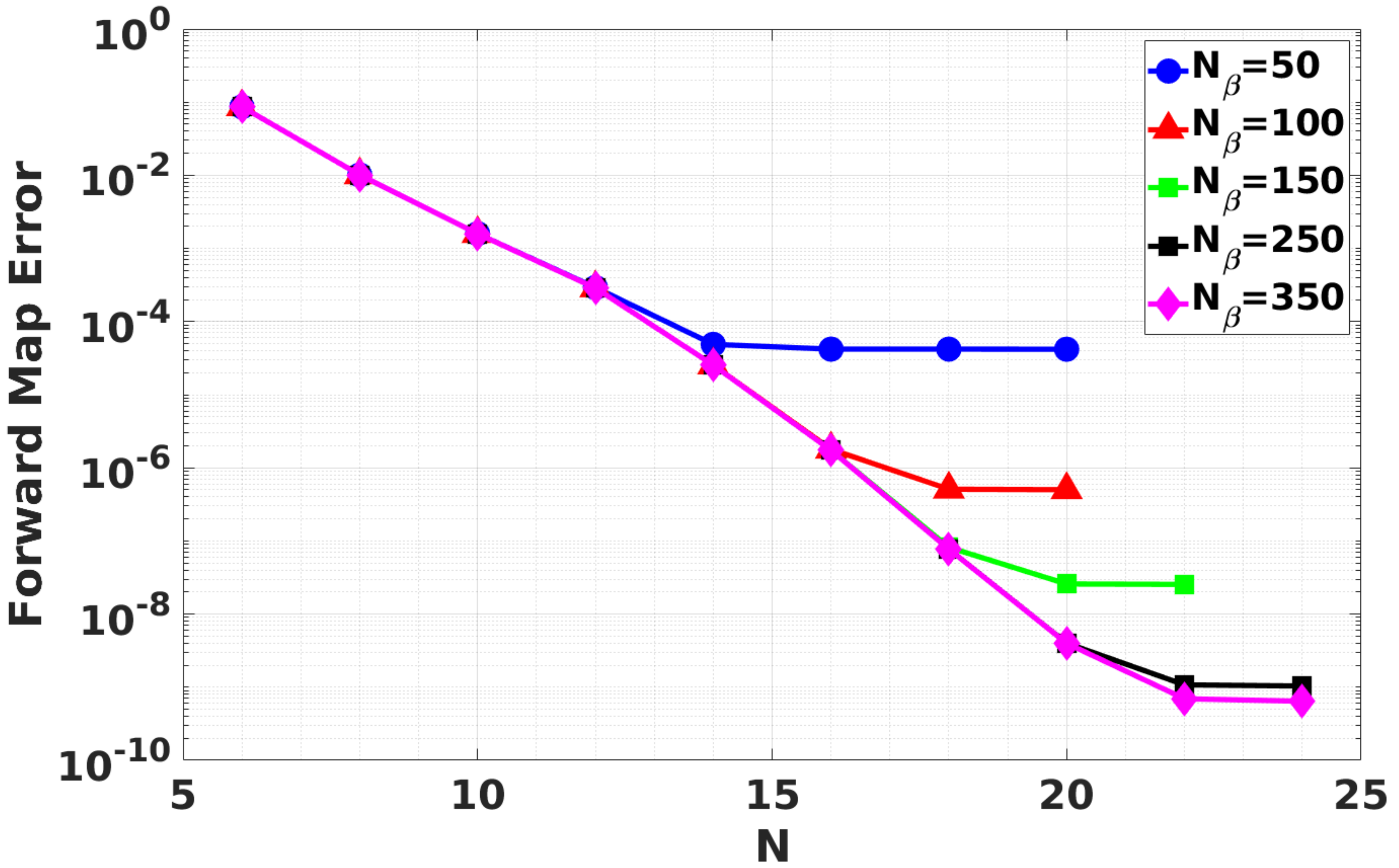}
}
\caption{(a) Forward mapping error with respect to $N$ for various choices of $N_\beta$ on a PEC sphere ($D=2\lambda$) using the MFIE formulation. (b) Forward mapping error on a dielectric sphere ($D=2\lambda_e, \epsilon_e=1.0, \epsilon_d=2.0$) using the N-M\"uller formulation.}
\label{fig:fwd_map_error}
\end{figure}
Fig.~\ref{fig:fwd_map_error} plots the forward mapping error (i.e. the error in the action of the integral operators when applied to a fixed reference current density) on a $2\lambda_e$ diameter sphere geometry for both the PEC and dielectric cases versus $N$ for various different choices of $N_\beta$. In the dielectric case, the exterior $\epsilon_e=1.0$ and the interior $\epsilon_d=2.0$. The Mie series solution due to an incident plane wave is used as the reference solution~\cite{harrington2001time}. As can be seen, depending on the desired accuracy, it is important to choose $N_\beta$ judiciously such that it does not limit the overall solution accuracy. Increasing $N_\beta$ does not increase the number of unknowns (controlled by $N$); however, it can significantly increase the amount of time required to precompute the singular and near-singular interactions.

\subsection{PEC Scattering: MFIE Formulation}
In this section, we test the proposed approach for the MFIE formulation by computing scattered fields from three PEC objects: two spheres of diameters $1.2\lambda$ and $4\lambda$ and a cube with side length $1.2\lambda$. All three objects are parameterized by using 6 patches, and each patch is discretized with the same number of points per patch per dimension $N=N_u=N_v$. Thus the total number of unknowns per problem is $Q=2\times6 \times N^2$. The spheres are illuminated by the same plane wave source, $\mathbf{E}^{\text{inc}} = \exp\left(-ikz\right)\mathbf{\hat{x}}$. Since a closed-form solution does not exist for scattering from a cube, we use an electric dipole excitation, $\mathbf{H}^{\text{inc}}(\mathbf{r})=-\nabla\times\left\{G(\mathbf{r}, \mathbf{r}')\mathbf{p}\right\}$, placed at position $\mathbf{r}'=(0.06\lambda,0.06\lambda,0.06\lambda)$ inside the cube with polarization $\mathbf{p}=(1,1,1)$. This allows us to determine convergence of the numerical solution since the scattered electric field must cancel the incident field outside the cube, and thus: $\mathbf{H}^\text{scat}(\mathbf{r}) = \nabla \times \left\{G(\mathbf{r},\mathbf{r}')\mathbf{p}\right\}$ for points $r$ outside of the cube. Note that in this case, the density solutions do not have a singularity at the cube edges, resulting in a similar convergence rate as in the case of the sphere. The results for the sphere cases are compared against the analytical Mie series solutions.

\begin{figure}[t]
\subfloat[][]{
\includegraphics[height=40mm]{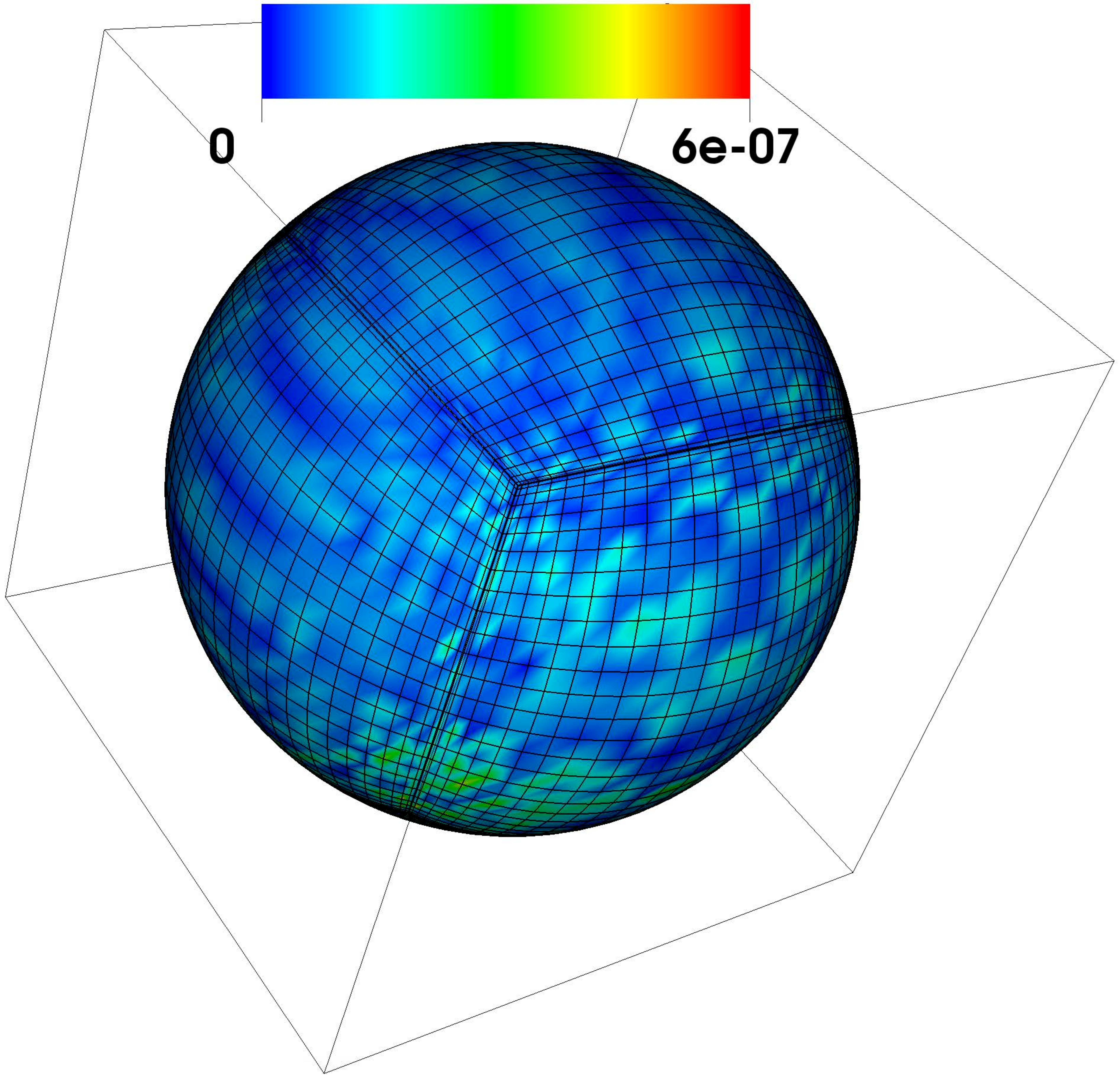}
}
\subfloat[][]{
\includegraphics[height=40mm]{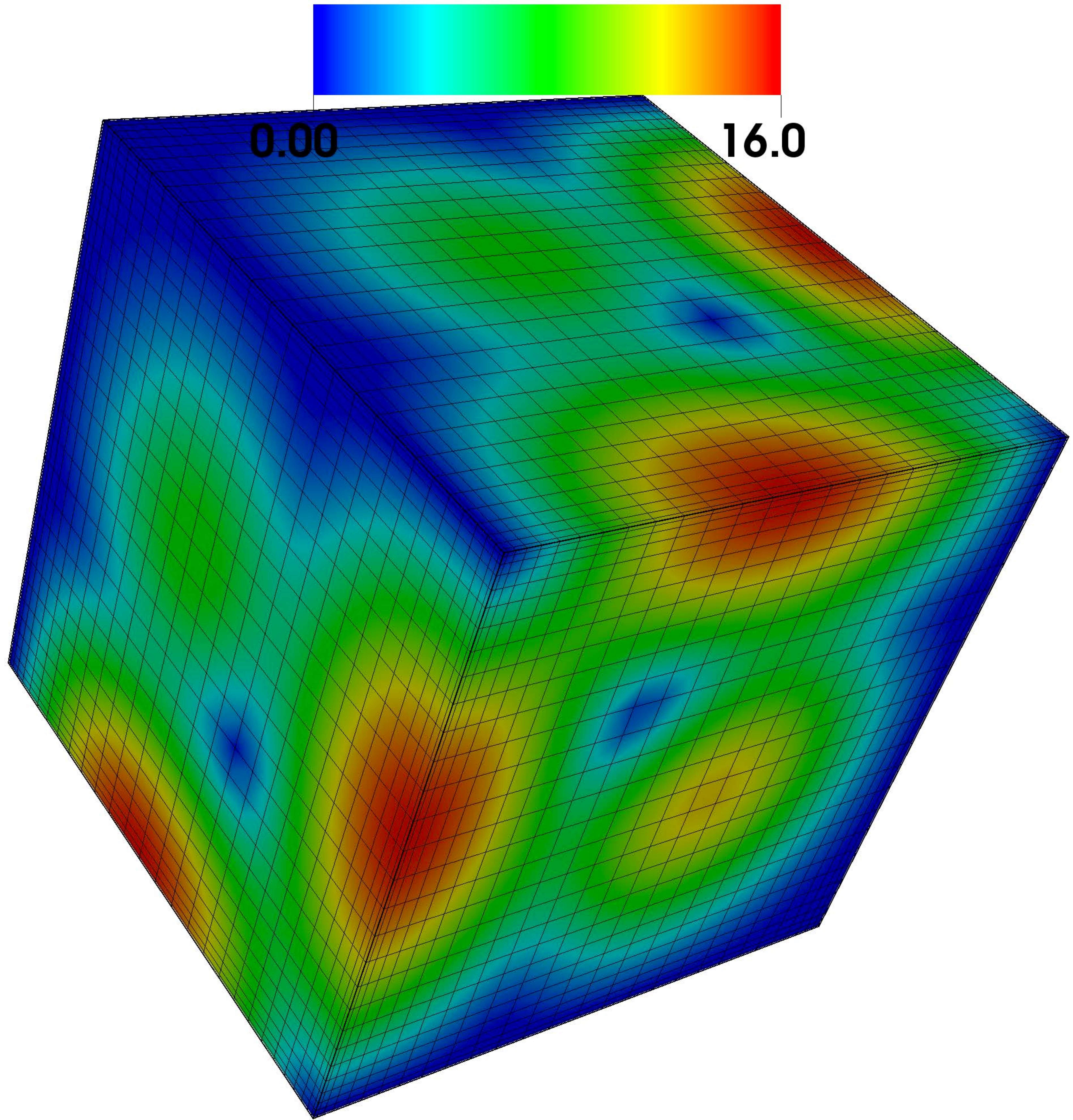}
}
\caption{(a) Error in the surface current distribution of a $4\lambda$ diameter sphere. The worst error is $5.7\times10^{-7}$. (c) Surface current distribution on a $1.2\lambda$ edge length cube.}
\label{fig:surfJ}
\end{figure}

Fig.~\ref{fig:surfJ}(a) shows the error in the surface density between the computed and analytical solution on the $4\lambda$ sphere for $N=26$. As can be seen, the numerical solution differs from the exact solution by less than $5.7\times10^{-7}$ at every point on the sphere. Fig.~\ref{fig:surfJ}(b) plots the computed surface current distribution on the cube resulting from the internal dipole source. 

\begin{figure}[ht]
    \centering
\includegraphics[width=\columnwidth]{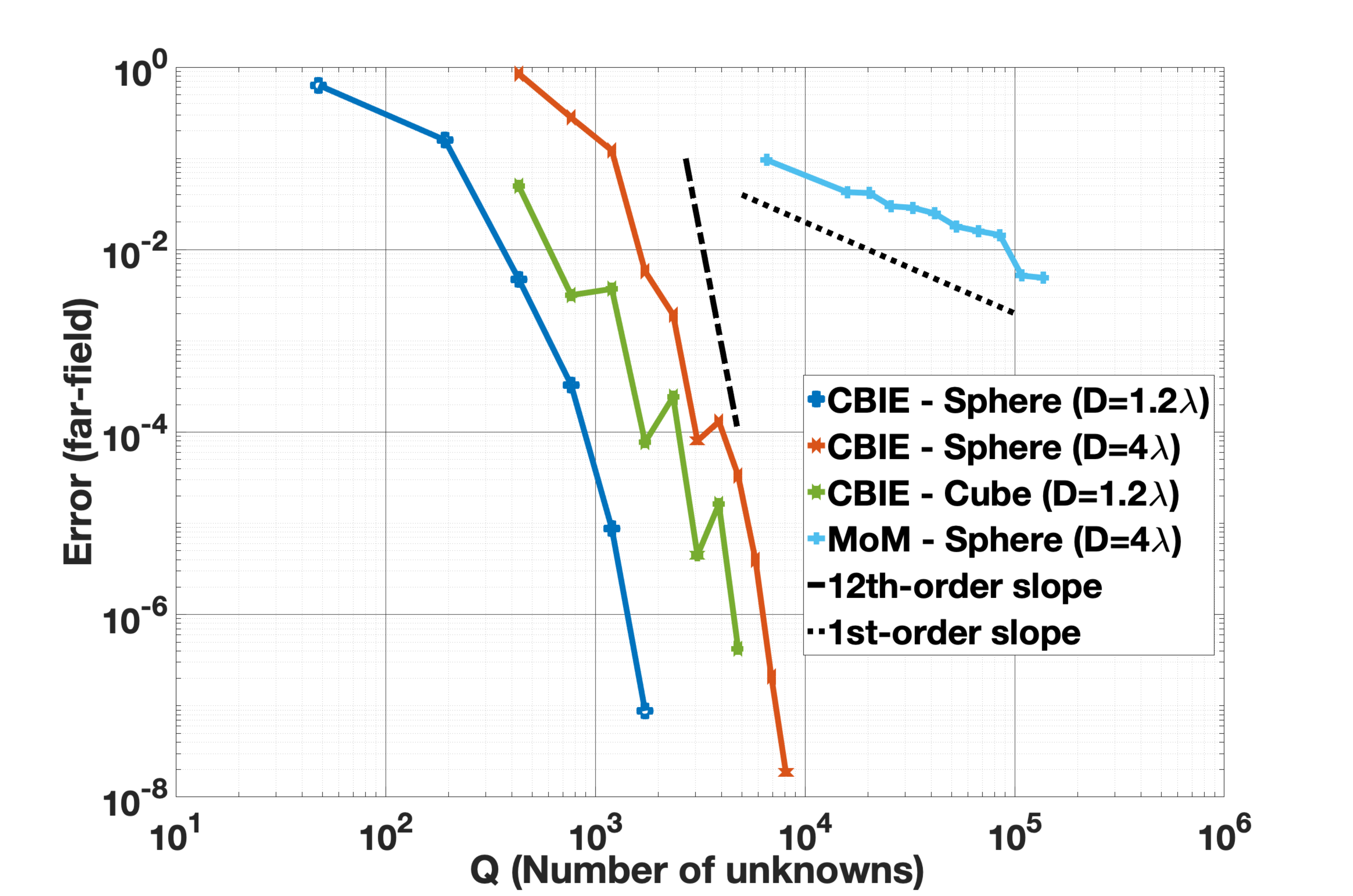}
\caption{Far-field error for the three scatterer examples vs. the number of unknowns. The performance of a commercial MoM RWG-based solver is also plotted for the sphere case with $D=4\lambda$ for comparison. $1^{\text{st}}$ and $12^{\text{th}}$ order asymptotes are drawn for reference.}
\label{fig:cbie_conv}
\end{figure}

Fig.~\ref{fig:cbie_conv} plots the error of the CBIE method vs. the number of unknowns ($Q$) used to discretize each scatterer. As a comparison, the convergence of a commercial MoM RWG-based solver for the $4\lambda$ sphere case is also plotted. For reference, $1^{\text{st}}$ and $12^{\text{th}}$ order slopes are drawn in dashed lines. As can be seen, the MoM solver only approaches first order convergence, requires a much finer discretization than the proposed CBIE method, and even for a very high resolution mesh barely exceeds two digits of accuracy. In contrast, the CBIE method converges spectrally fast for all three examples, which makes it a significantly more accurate and efficient approach.

\subsection{Dielectric Scattering: N-M\"uller Formulation}
The scattered fields from two dielectric objects are computed to evaluate the performance of the CBIE method for the N-M\"uller formulation: a dielectric sphere of 2$\lambda_e$ diameter with permittivity $\epsilon_d = 2\epsilon_e$ and a dielectric cube of 2$\lambda_e$ side length with permittivity $\epsilon_d = 2\epsilon_e$, where the $\lambda_e=2\pi /k_e$ is the wavelength corresponding the background exterior medium which is set to free-space for all problems considered here ($\epsilon_e = \epsilon_0$). The magnetic permeability for both objects is also set to the vacuum permeability: $\mu_d = \mu_e = \mu_0$. The surfaces of the objects are discretized in the same manner as for the MFIE formulation, which results in $Q=2\times2\times6 \times N^2$ unknowns. They are both illuminated by a plane wave excitation $\mathbf{E}^{\text{inc}} = \exp\left(-ikz\right)\mathbf{\hat{x}}$. The results are compared against the Mie series analytical solution for the dielectric sphere~\cite{harrington2001time} and against a highly refined numerical solution for the dielectric cube since an analytical solution does not exist. 

\begin{figure}[t]
\centering
\subfloat[][]{
    \includegraphics[width=37mm]{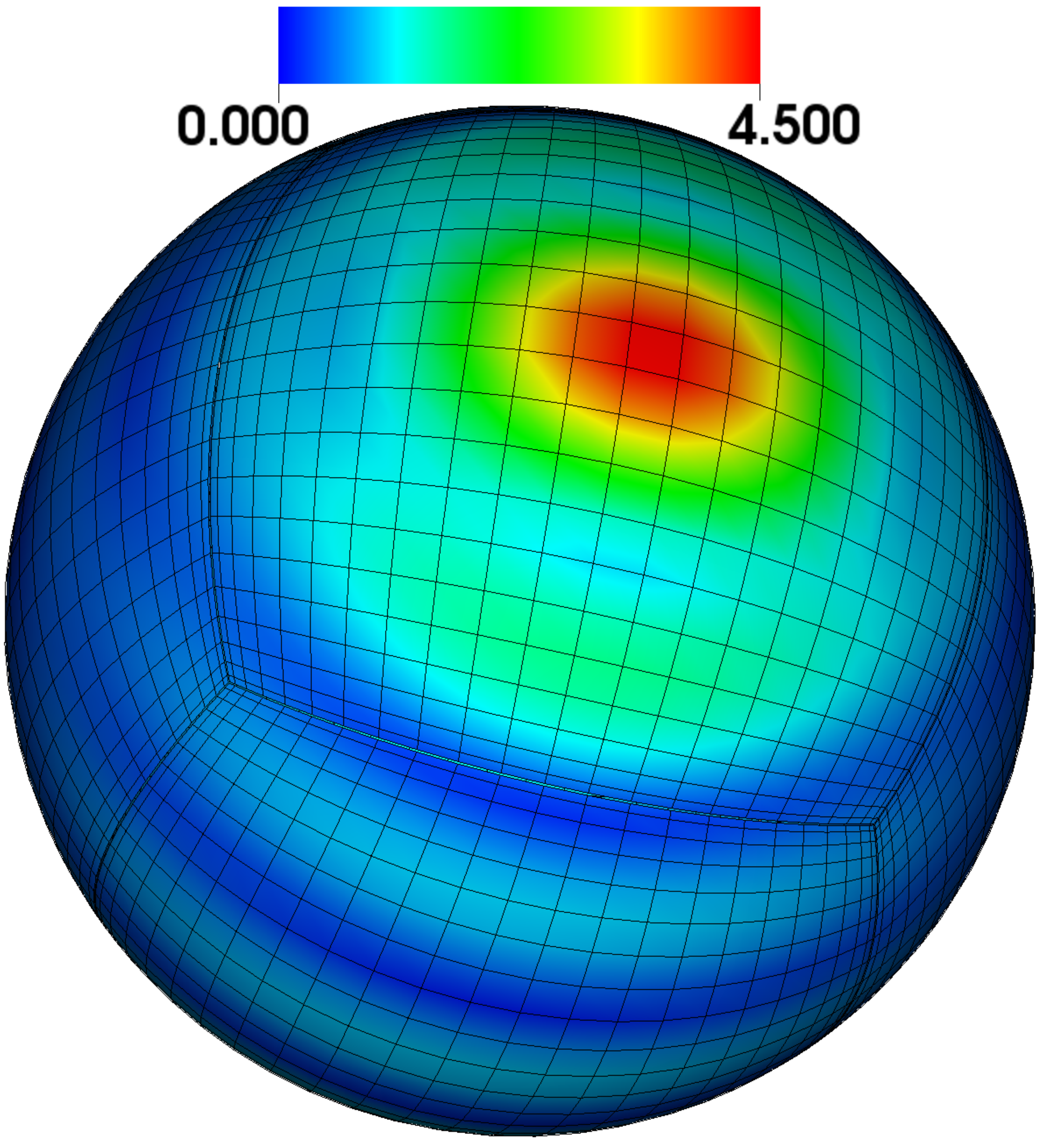}
}
\subfloat[][]{
    \includegraphics[width=37mm]{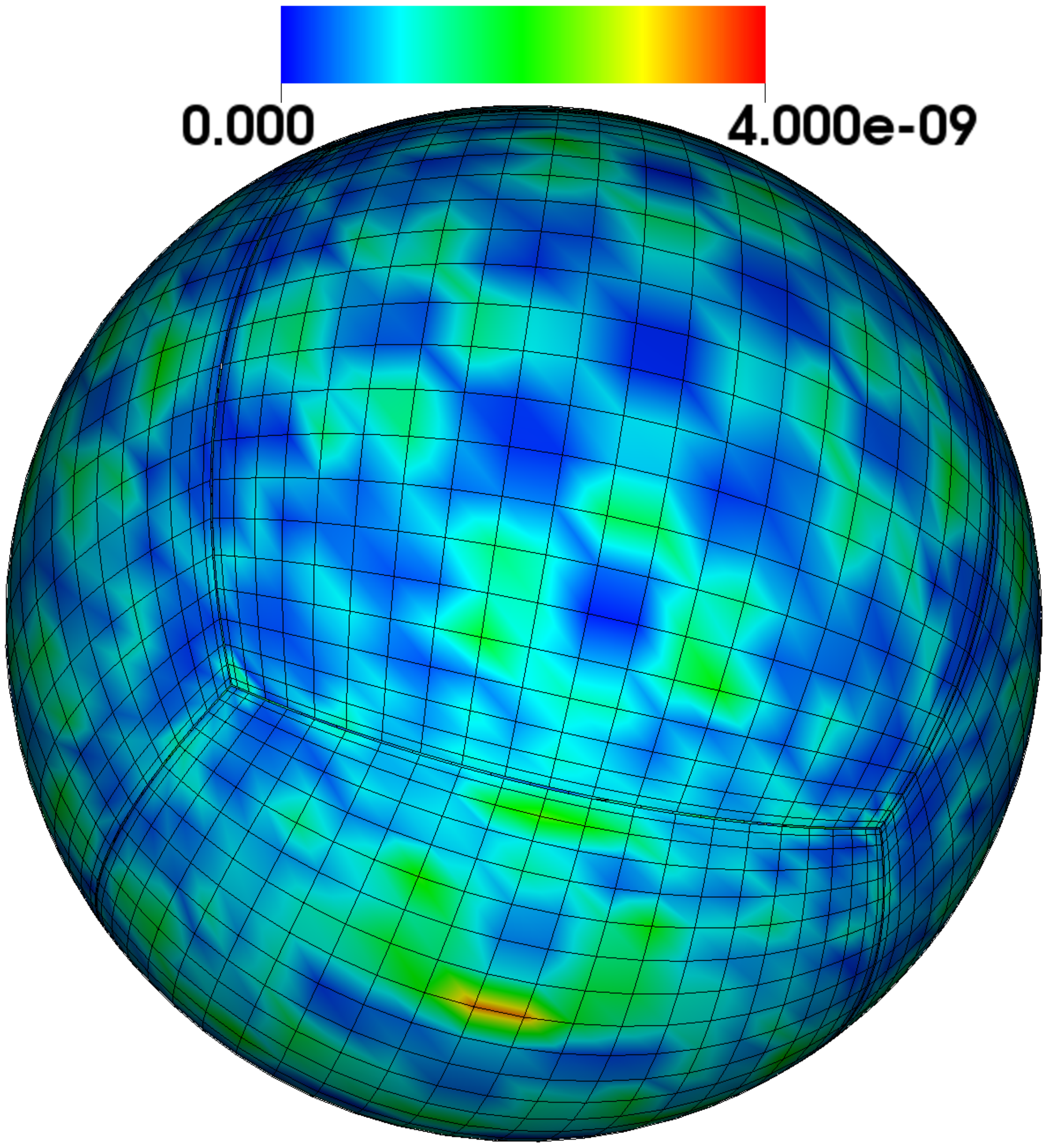}
} 
\\
\subfloat[][]{
    \includegraphics[width=41mm]{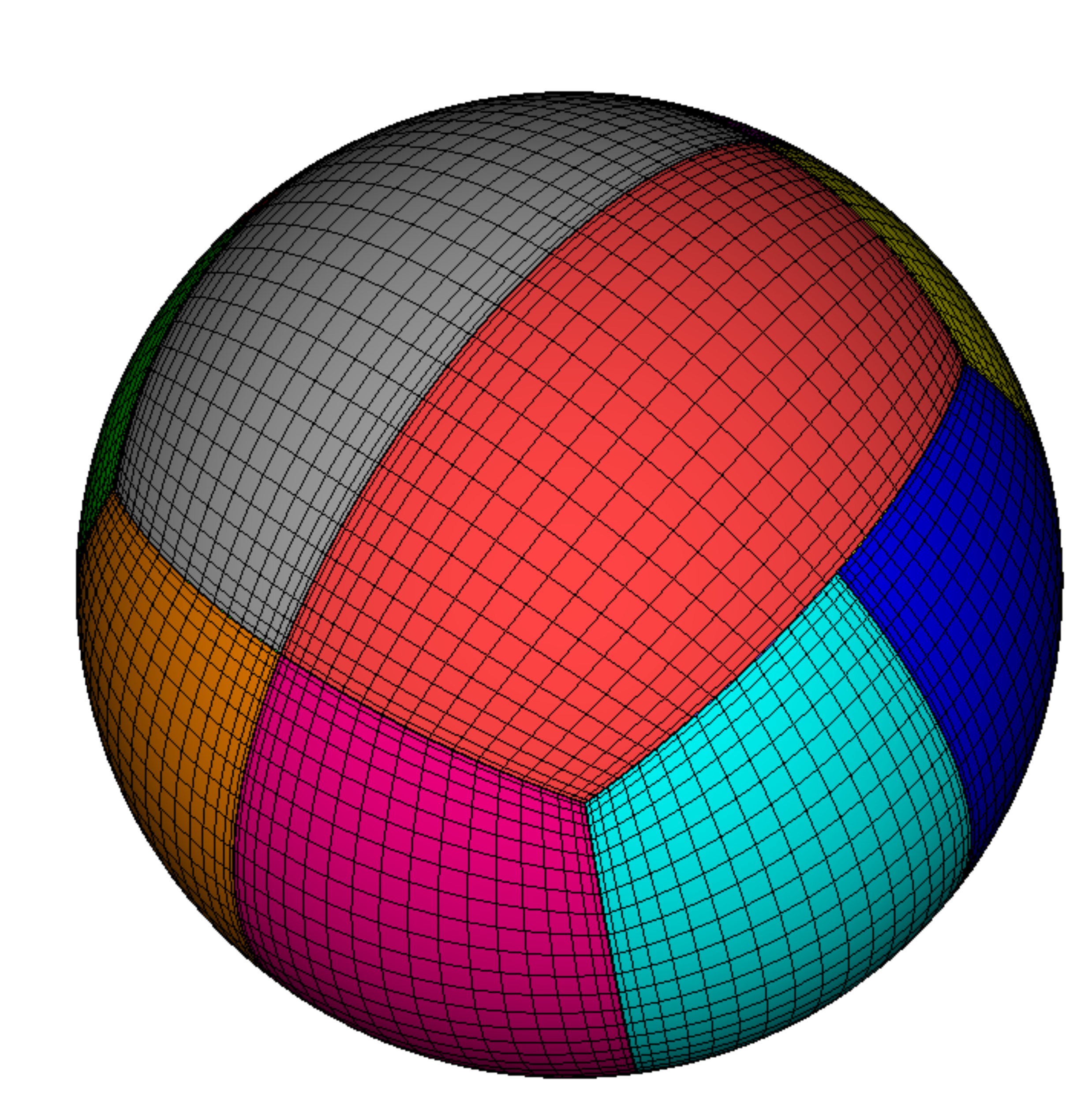}
}
    \subfloat[][]{
\includegraphics[width=41mm]{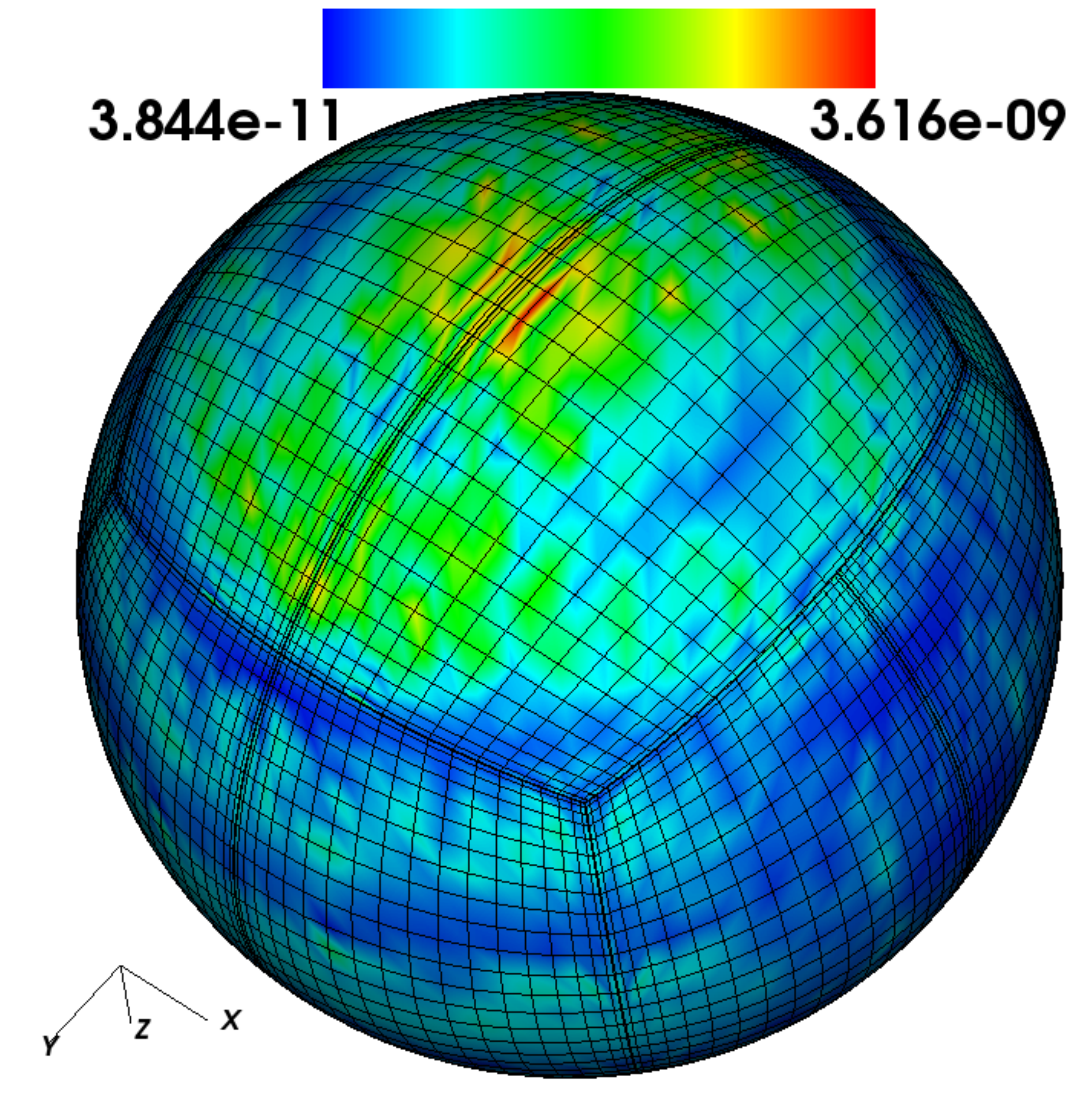}
}
\caption{(a) Surface M distribution on $2\lambda_e$ diameter dielectric sphere with $\epsilon_d=2\epsilon_e$. (b) Error of surface M distribution. Max error: $3.5\times10^{-9}$. (c) Patch configuration for a defective mesh. (d) Error of surface M distribution for the defective mesh. Max error: $3.6 \times 10^{-9}$.} \label{fig:surfJM} 
\end{figure}

Fig.~\ref{fig:surfJM}(a) shows the absolute value of the magnetic (M) current density distribution on the surface of the $2\lambda_e$ sphere for $N=24$ and Fig.~\ref{fig:surfJM}(b) shows the error difference of the computed current density distribution with the Mie Series solution. Fig.~\ref{fig:surfJM}(c)  show the patch configuration for a ``defective'' mesh, i.e. one or more patch edges are only partially shared by another quadrilateral patch, and Fig.~\ref{fig:surfJM}(d) shows the corresponding pointwise error on the magnetic current density using 30 points per patch along the largest dimension, and 20 for the smallest patch dimension. This example demonstrates one of the advantages of  using a Nystr\"om method and quadrilateral patches with an open-grid quadrature.

\begin{figure}[ht]
    \centering
\includegraphics[width=\columnwidth]{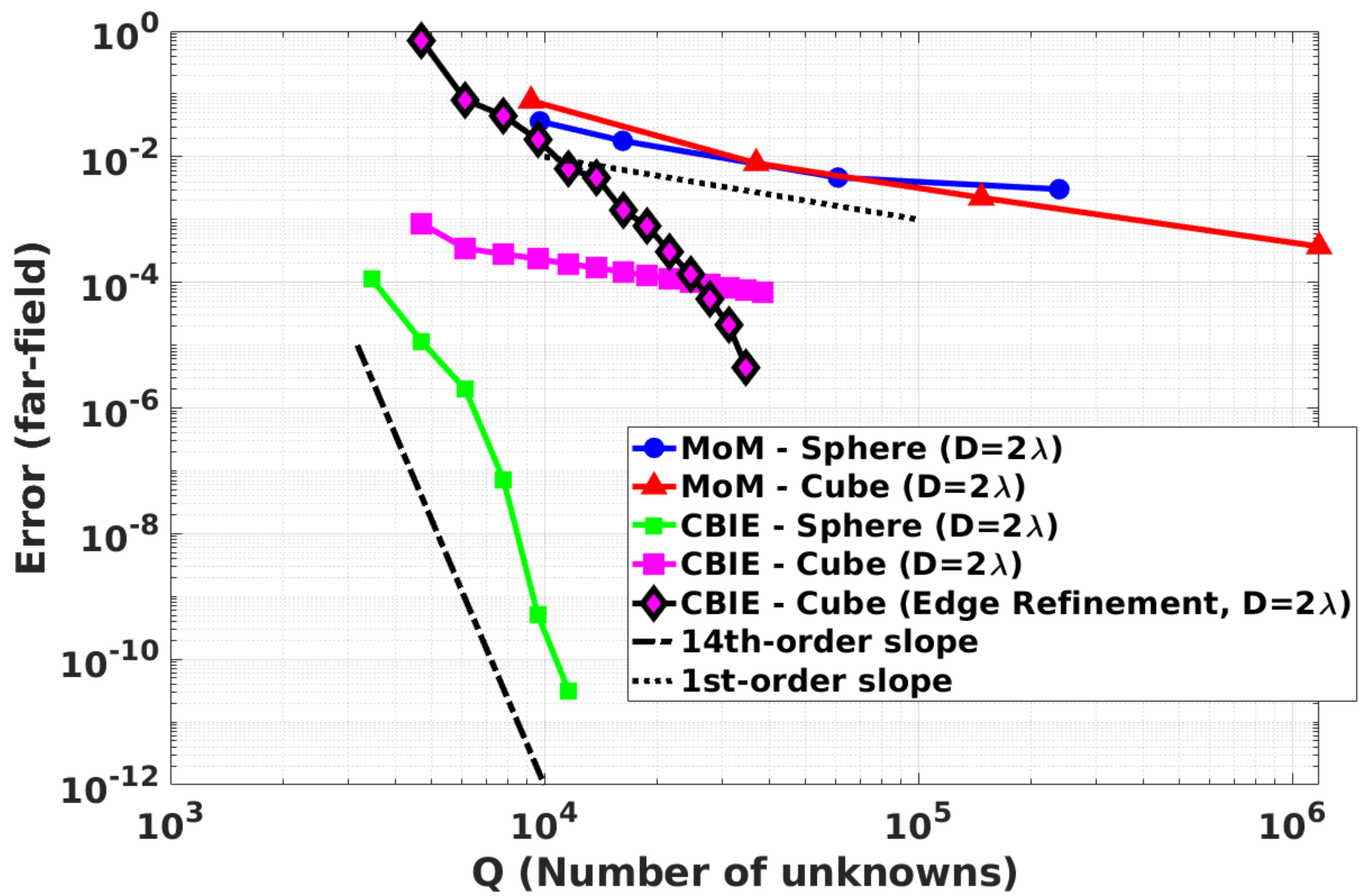}
\caption{Convergence of far-field error for the two dielectric scatterer examples vs. number of unknowns. Convergence for the dielectric cube using edge refinement is also plotted. Performance of commercial MoM RWG-based solver is shown for comparison. $2^{\text{nd}}$ and $14^{\text{th}}$ order asymptotes are drawn for reference.}
\label{fig:dielec_conv}
\end{figure}

Fig.~\ref{fig:dielec_conv} plots the error of the CBIE method vs. the number of unknowns ($Q$) used to discretize each scatterer. As expected, the convergence for the cube is considerably worse than that of the sphere due to the edge and corner singularities in the current densities. The convergence rate can be recovered, however, by using the same edge refinement approach proposed in~\cite{bruno2018chebyshev} which clusters unknowns near the edges to better resolve the singularities. This improvement can be seen in the edge refined curve plotted in Fig.~\ref{fig:dielec_conv}.  As a comparison, the convergence of a commercial MoM RWG-based solver for the both objects is also plotted. For reference, $1^{\text{st}}$ and $14^{\text{th}}$ order slopes are drawn in dashed lines. As with the PEC case, the MoM solver only approaches first order convergence and requires a much finer discretization than the proposed CBIE method due to the linear basis functions and flat triangular discretization used to represent the geometry.

For all of the previous examples, a direct linear algebra solver was used to obtained the density solutions, and in Table~\ref{tab:cond} we show the condition number for various discretizations on both the MFIE and N-M\"{u}ller systems. Indeed, these condition numbers compare well to other high-order methods, including the mixed-order basis, locally corrected method from~\cite[Tables~I-III]{Gedney2004}. To further show the ease at which this methodology can be incorporated with an iterative solver that does not require explicit formation of the system matrices, we present in Table~\ref{tab:h} the convergence of the method while performing ``h-refinement''---increasing the number of patches $M$ while keeping the number of points per patch per dimension $N$ constant---for a dielectric sphere with the same parameters as in Fig.~\ref{fig:surfJM}. In Table~\ref{tab:h} we also show the total number of unknowns $Q$ and the number of iterations needed by the iterative method GMRES. The tolerance for GMRES was set to $10^{-5}$ for $N=6$ and $N=8$, and $10^{-7}$ for $N=10$. All timings correspond to simulations using 6 cores of an Intel i9-9900KF running at 4.7GHz.

\begin{figure}[t]
    \centering
    \subfloat[][]{
    \includegraphics[width=40mm]{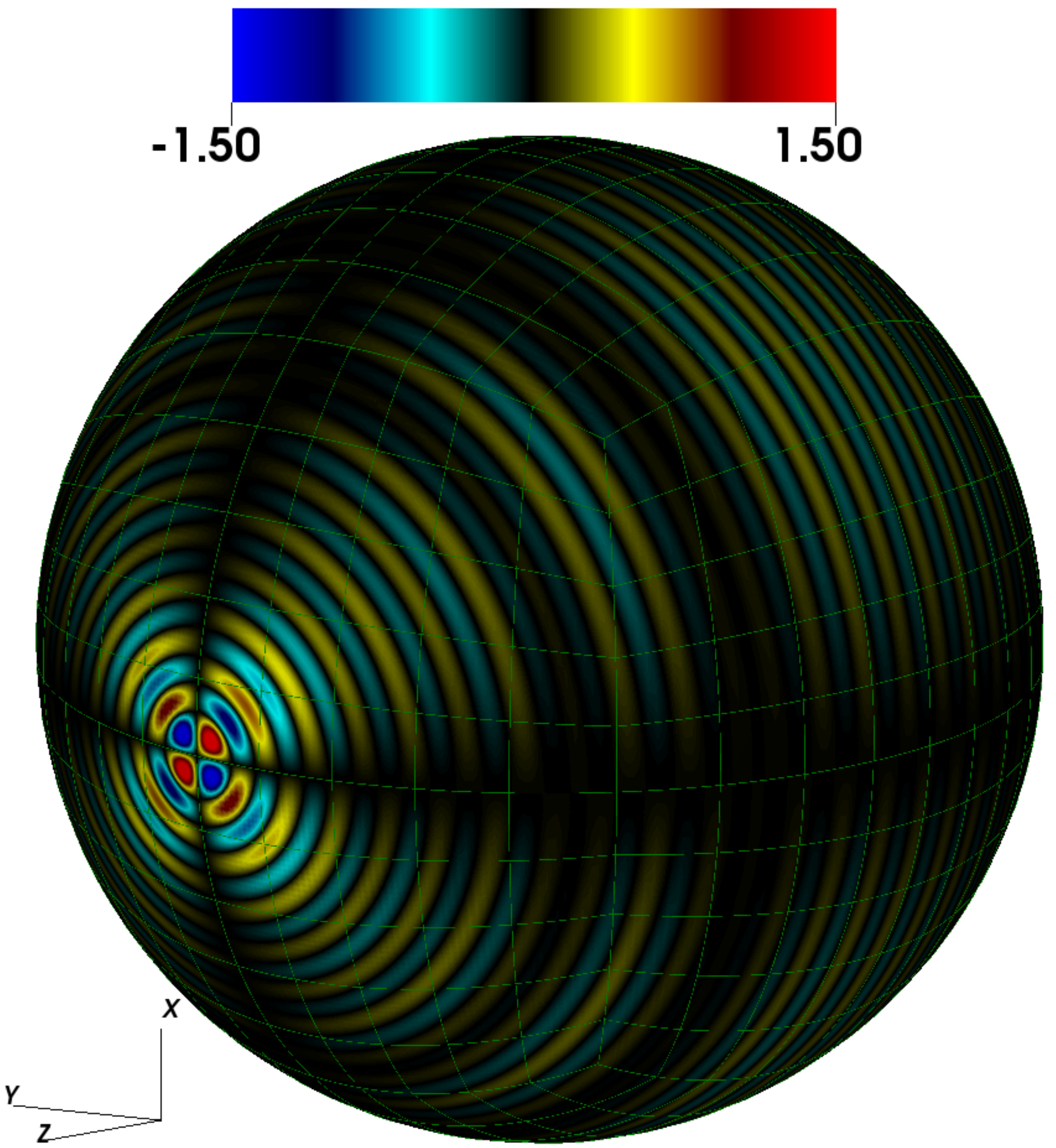}
    } 
    \subfloat[][]{
    \includegraphics[width=40mm]{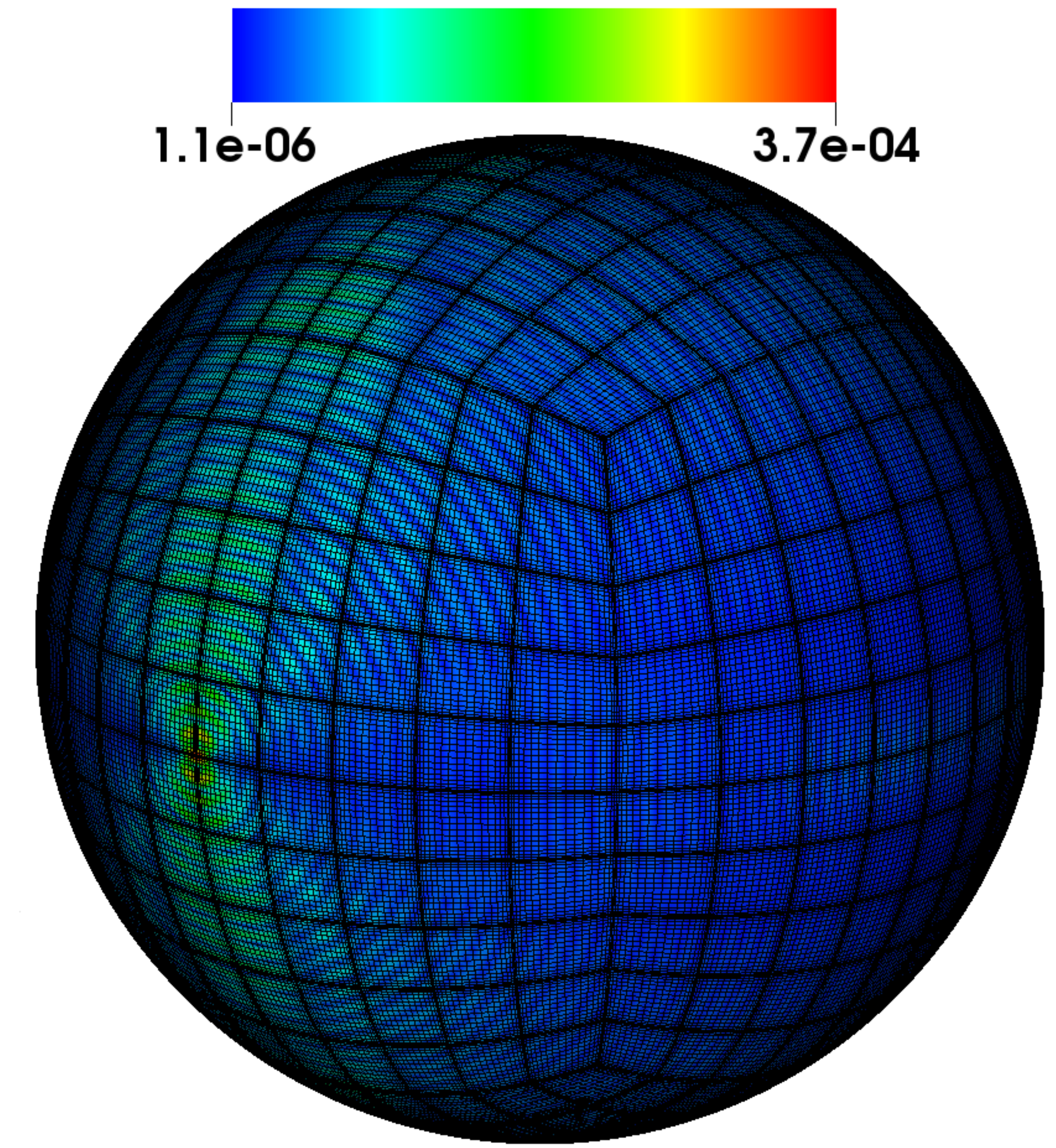}
    } \\
    \subfloat[][]{
    \includegraphics[width=78mm]{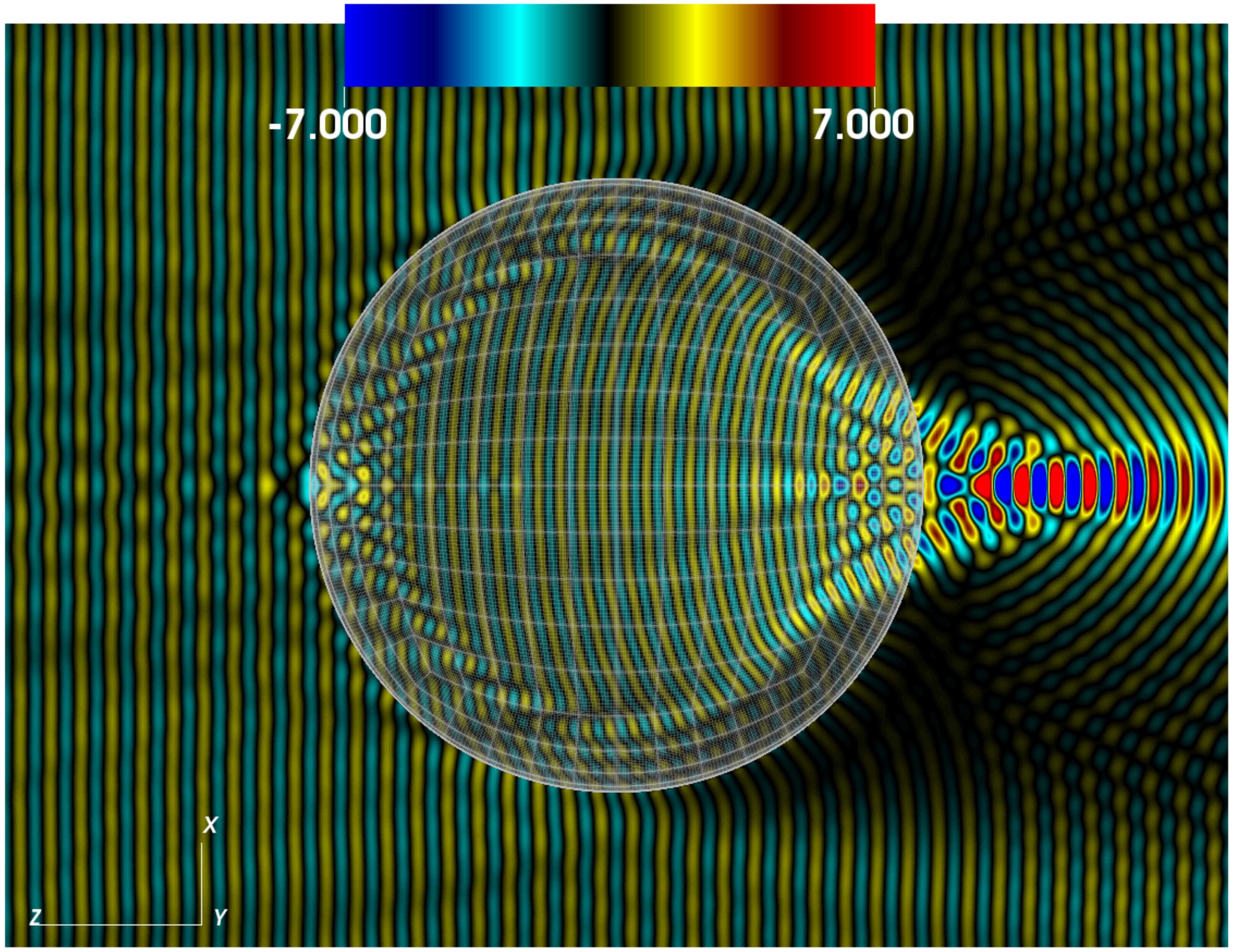}
    } \\
    \subfloat[][]{
    \includegraphics[width=78mm]{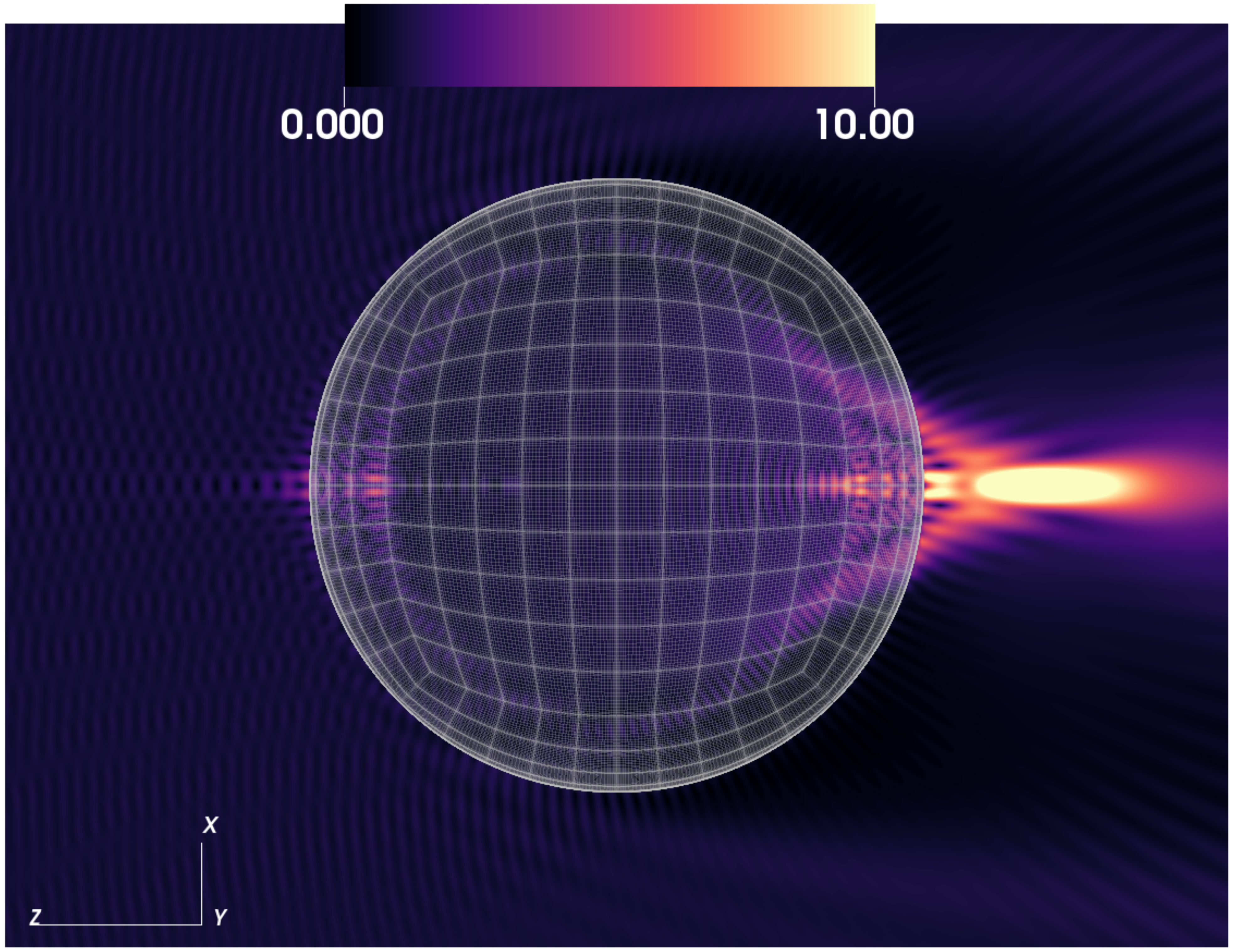}
    } 
    \caption{(a) Real part of the $x$-component of the current density $\mathbf{M}$ for scattering by a sphere of diameter $20\lambda_e$ (=$28.3\lambda_d$). (b) Pointwise error for a discretization consisting of 600 patches each with $18$ points per dimension. In (c) and (d) the real part (of the $x-$component) and the absolute value of the electric field are shown, respectively.}
    \label{fig:largeSph}
\end{figure}

In Fig.~\ref{fig:largeSph} we show a simulation of a large dielectric sphere with diameter $20 \lambda_e$ ($=28.3\lambda_d$) using a discretization of 600 patches and $N=18$ for a total of 194,400 discretization points and 777,600 unknowns. Fig.~\ref{fig:largeSph}(a) shows the real part of the $x$-component of the density $\mathbf{M}$. Fig.~\ref{fig:largeSph}(b) shows the pointwise error in the density $\mathbf{M}$, with a maximum value of $3.7 \times 10^{-4}$ (for a GMRES tolerance of $10^{-5}$). The real part of the $x$-component of the electric field, and the absolute value of the total electric field are shown in Fig.~\ref{fig:largeSph}(c)~and~(d) respectively.

\begin{table}
    \centering
    \caption{Condition numbers for the discretized MFIE and N-M\"{u}ller systems. The spheres and cubes are of diameter and side lengths of $2\lambda$, respectively, with $\epsilon_d = 2\epsilon_e$ for the dielectric case. }
    \label{tab:cond}
    \begin{tabular}{| c | c | c | c | c |}
        \hline
         & \multicolumn{2}{c|}{MFIE} & \multicolumn{2}{c|}{ N-M\"{u}ller} \\
         \hline
        $N$ & Sphere  & Cube & Sphere  & Cube \\
        \hline   \hline
        8  & 13.76 & 28.46 & 46.29 & 45.15 \\
        10 & 13.81 & 29.95 & 49.99 & 47.80 \\
        12 & 13.82 & 31.18 & 53.87 & 49.44 \\
        14 & 13.82 & 32.35 & 57.61 & 51.51 \\
        16 & 13.82 & 33.35 & 61.27 & 53.48 \\
        18 & 13.82 & 34.25 & 64.89 & 55.43 \\
        20 & 13.82 & 35.06 & 68.48 & 57.35 \\
        \hline
    \end{tabular}
\end{table}

\begin{table}
    \centering
    \caption{Convergence by increasing the number of patches and keeping the degree of the expansion $N$ constant for a dielectric sphere of diameter $2 \lambda_e$. The times for the precomputations and for finding the current density solutions via GMRES are all in seconds.}
    \label{tab:h}
    \begin{tabular}{| c | c | c | c | c | c | c |}
        \hline
        $N$ & $M$ & $Q$ & GMRES & Prec. (s) & Solve (s) & Error \\
        \hline         \hline
        6 & 24  & 3,456  & 24 & 0.7  & 0.3 & 4.2$\times 10^{-2}$ \\
        6 & 54  & 7,776  & 20 & 1.8  & 1.2 & 7.2$\times 10^{-3}$ \\
        6 & 96  & 13,824 & 20 & 4.0  & 3.5 & 2.2$\times 10^{-3}$ \\
        6 & 150 & 21,600 & 20 & 7.3  & 8.6 & 8.6$\times 10^{-4}$ \\
        6 & 216 & 31,104 & 20 & 11.6 & 17.3 & 4.2$\times 10^{-4}$ \\
        \hline
        8 & 24  & 6,144  & 20 & 1.3  & 0.6  & 1.9$\times 10^{-3}$ \\
        8 & 54  & 13,824 & 20 & 3.8  & 3.3  & 2.1$\times 10^{-4}$ \\
        8 & 96  & 24,576 & 20 & 7.7  & 10.5 & 3.5$\times 10^{-5}$ \\
        8 & 150 & 38,400 & 20 & 13.8 & 25.4 & 8.9$\times 10^{-6}$ \\
        8 & 216 & 55,296 & 20 & 23.8 & 52.4 & 3.6$\times 10^{-6}$ \\
        \hline
        10 & 24  & 9,600 & 24 & 2.2 & 1.9  & 5.7$\times 10^{-5}$ \\
        10 & 54  & 21,600 & 24 & 6.5 & 9.3  & 5.0$\times 10^{-6}$ \\
        10 & 96  & 38,400 & 24 & 12.8 & 29.9 & 8.9$\times 10^{-7}$ \\
        10 & 150 & 60,000 & 24 & 25.6 & 73.5 & 2.2$\times 10^{-7}$ \\
        10 & 216 & 86,400 & 24 & 40.1 & 152.0 & 7.5$\times 10^{-8}$ \\
        \hline
    \end{tabular}
\end{table}

\subsection{Scattering from Complex NURBS CAD Models}
\begin{figure}[t]
\centering
\subfloat[][]{
\includegraphics[width=18mm]{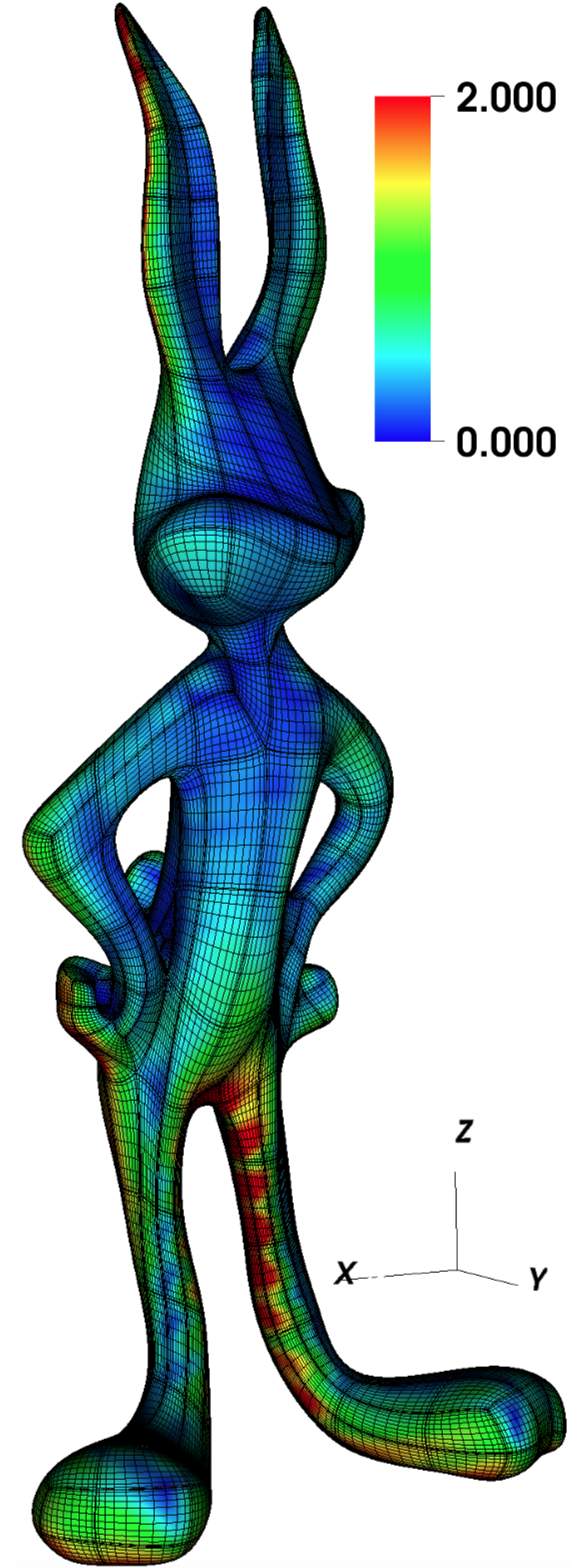}
} 
\subfloat[][]{
\includegraphics[width=65mm]{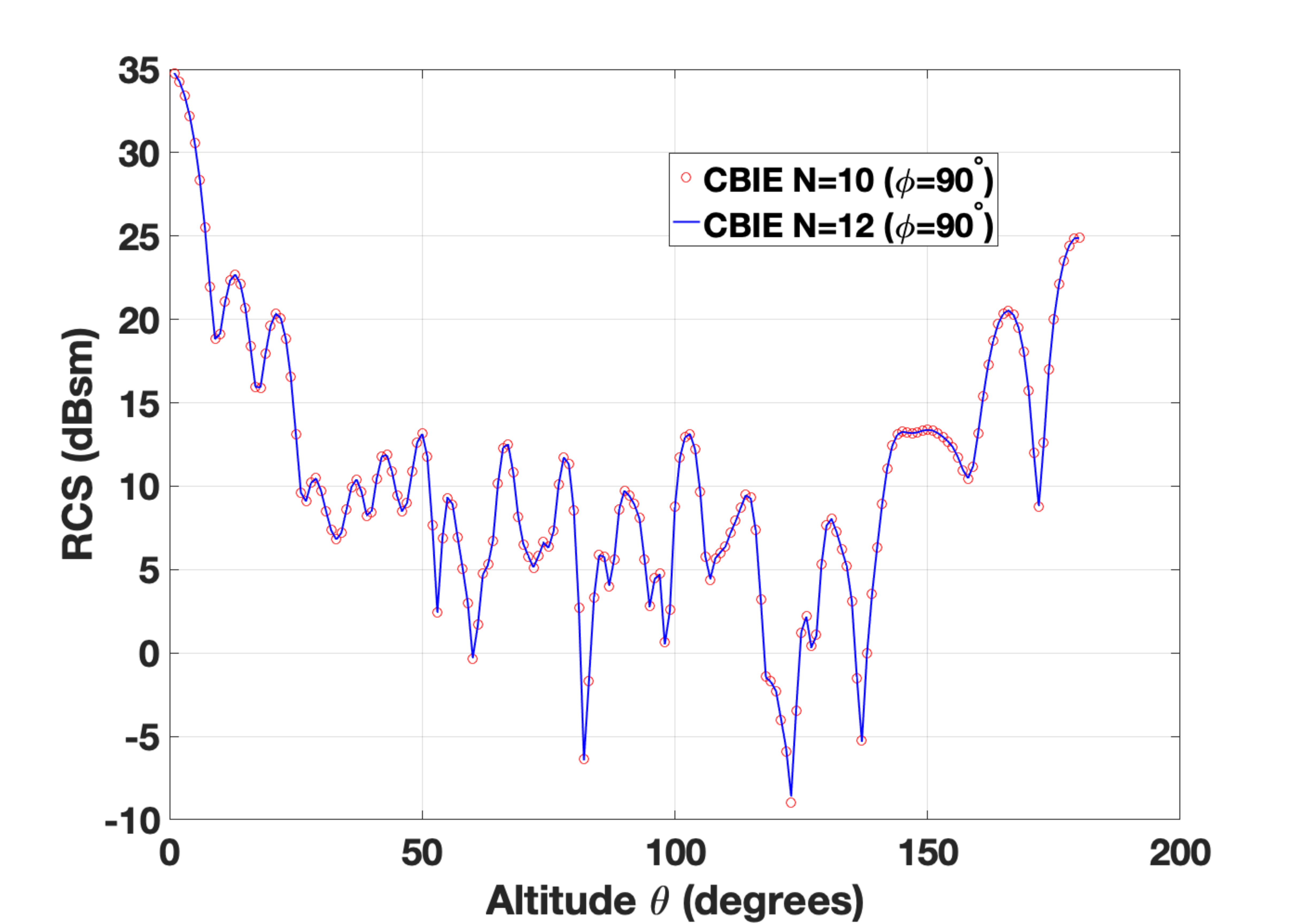}
} 

\caption{(a) Surface electric current density induced on $16\lambda$ tall PEC CAD humanoid bunny model by incident plane wave. The model consists of 402 curvilinear quadrilateral NURBS-parametrized patches. (b) RCS at $\phi=90^{\circ}$  corresponding to plane wave scattering for $N=10$ and $N=12$ Chebyshev points per patch discretizations.}
\label{fig:CAD_bugs}
\end{figure}

In order to demonstrate that the proposed approach can be readily used to solve scattering from complex CAD generated models with arbitrary curvature, we solve for the scattered fields from two different NURBS models freely available for download online~\cite{grabcad}. As in the previous examples, the incident excitation is an $x$-polarized plane wave propagating in the $+z$ direction. In the first example, we consider scattering off of a 16 wavelength tall humanoid bunny character. Fig.~\ref{fig:CAD_bugs}(a) shows the induced surface current density and Fig.~\ref{fig:CAD_bugs}(b) plots the RCS vs. $\theta$ at a $\phi=90^{\circ}$ angle for two different discretizations ($N=10$ and $N=12$ Chebyshev points per side per patch or 100 and 144 points per patch total respectively). The model is comprised of 402 curvilinear quadrilateral patches total and was directly imported from a standard CAD software without any special post-processing required~\cite{rhino_cite}. Despite the large size of the model, significant variation in curvature, and regions with sharp geometrical features (e.g., the ears), the match in the RCS for the two relatively coarse discretizations is excellent and they are almost indistinguishable from one another, varying less than $1\times10^{-4}$ from each other.  

\begin{figure}[ht]
\centering
\subfloat[][]{
\includegraphics[width=70mm]{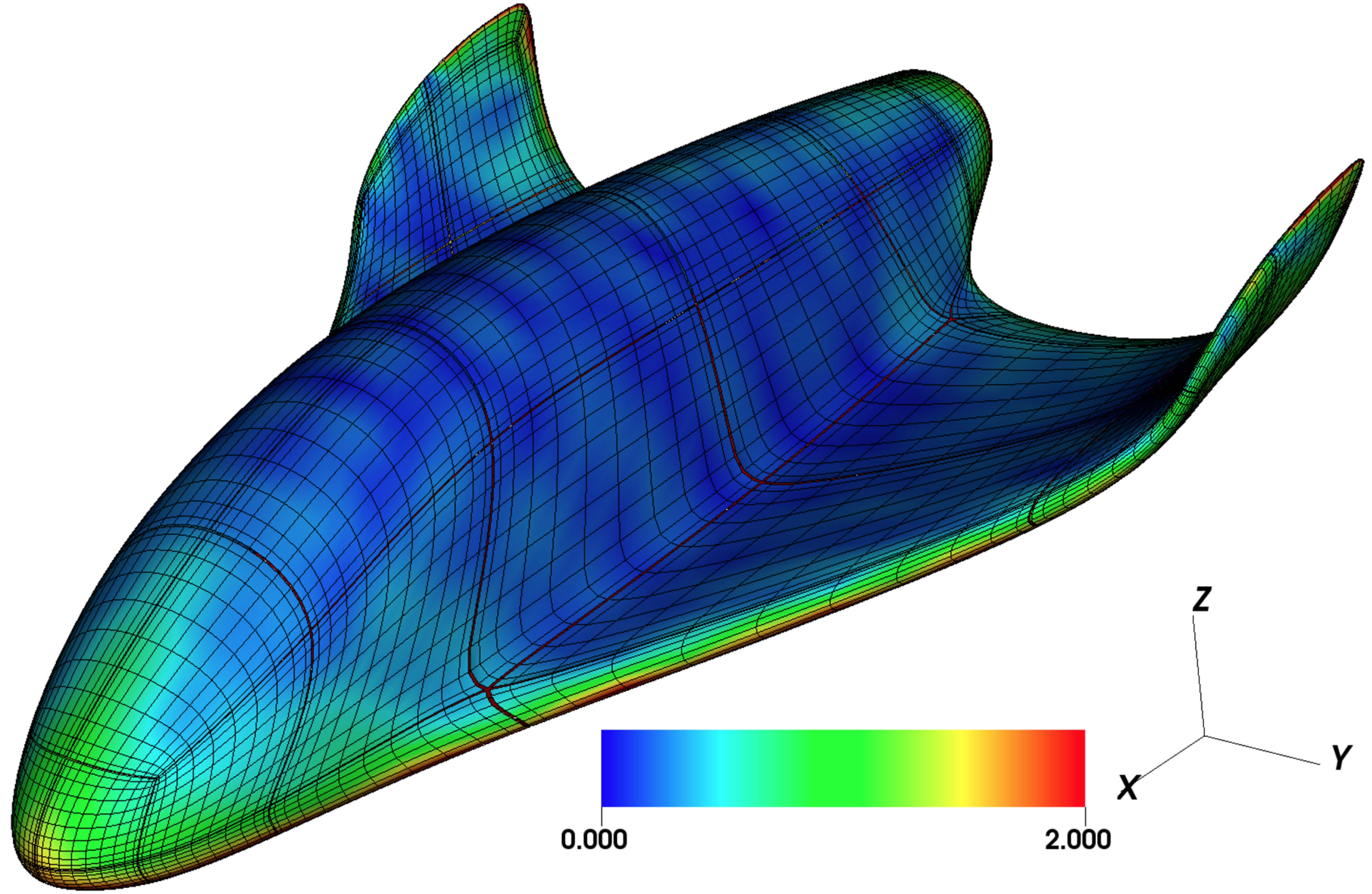}
}\\
\subfloat[][]{
\includegraphics[width=85mm]{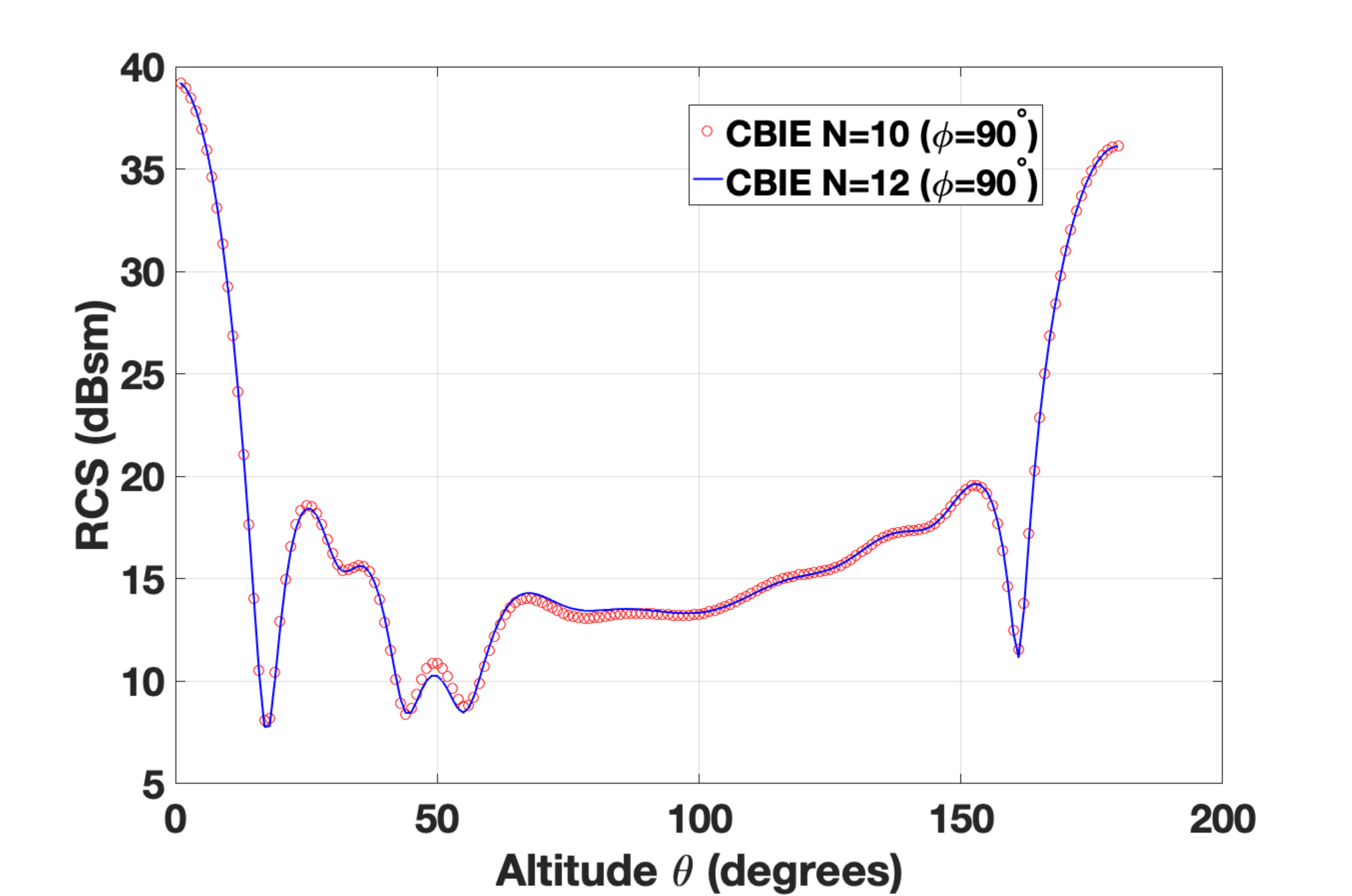}
} 
\caption{(a) Surface electric current density induced on 79 patch PEC glider CAD model by an incident plane wave. The glider spans 7.7 wavelengths from wing to wing. (b) RCS at $\phi=90^{\circ}$  corresponding to plane wave scattering for $N=10$ and $N=12$ Chebyshev points per patch discretizations.}
\label{fig:CAD_glider}
\end{figure}

For the second CAD model example, we computed scattering from a glider with a length of 7.7 wavelengths and a wingspan of 5.6 wavelengths from the end of one wing to the other. Fig.~\ref{fig:CAD_glider}(a) shows the induced surface current density and Fig.~\ref{fig:CAD_glider}(b) plots the RCS vs. $\theta$ at a $\phi=90^{\circ}$ angle for two different discretizations ($N=10$ and $N=12$ Chebyshev points per side). The glider is comprised of 79 curvilinear quadrilateral patches total. As before, the RCS curves resulting from the two different discretizations match very well and vary less than $2.5\times10^{-2}$ from each other.

\section{Conclusion}
This paper presents a high-order accurate Chebyshev-based Boundary Integral Equation (CBIE) approach for solving Maxwell's equations. The CBIE method is applied towards the discretization of the MFIE and the N-M\"uller formulation. The performance is evaluated by solving scattering from sphere and cube PEC/dielectric objects and comparing against analytical solutions as well as a commercial MoM-based solver. We have also demonstrated a couple examples of scattering from complex 3D CAD models which contain many intricate features and variations in curvature. The proposed method achieves spectral convergence on sufficiently smooth surfaces with respect to the number of unknowns, significantly reducing the number of unknowns required for a desired accuracy over low-order MoM approaches. Furthermore, the CBIE approach also converges well for geometries with edges and corners when an edge-refinement change of variables is utilized as demonstrated by the dielectric cube example. Current and future work involves applying the CBIE method in conjunction with the Windowed Green Function (WGF)\cite{bruno2017windowed} method towards the simulation and design of 3D waveguiding structures with unbounded boundaries for modeling nanophotonic devices~\cite{sideris2019ultrafast}, treating multi-material and composite objects~\cite{yla2005surface, perez2019planewave,perez2019planewave2}, and incorporating acceleration techniques such as the Fast Multiple Method~\cite{Greengard1987,engheta1992fast,Greengard1998,Gumerov2004} or FFT-based methods~\cite{Bruno2001, bleszynski1996aim}.~\nocite{HPV:VisIt}

\bibliographystyle{IEEEtran}
\bibliography{IEEEabrv,bibliography} 

\newpage

\begin{IEEEbiography}[{\includegraphics[width=1in,height=1.25in,clip,keepaspectratio]{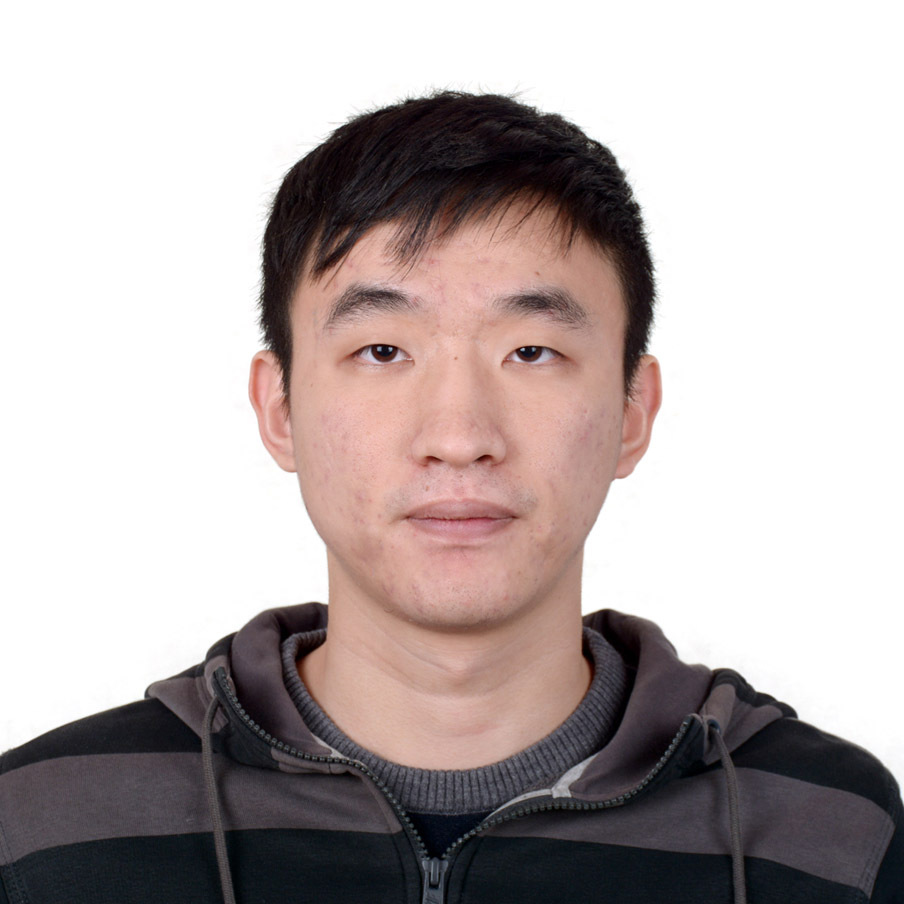}}]{Jin Hu}
    received the B.Eng. degree in electronic information engineering from the University of Science and Technology of China, Hefei, China in 2016 and the M.S. degree in electrical engineering from the University of Southern California, Los Angeles, CA, USA in 2019. He is currently pursuing the PhD degree with the Ming Hsieh Department of Electrical and Computer Engineering, the University of Southern California, Los Angeles, CA, USA. His research interests include the boundary integral equation methods for electromagnetics scattering analysis and its applications in simulation and design of nanophotonic devices.
\end{IEEEbiography}

\vspace{-4.4in}

\begin{IEEEbiography}[{\includegraphics[width=1in,height=1.25in,clip,keepaspectratio]{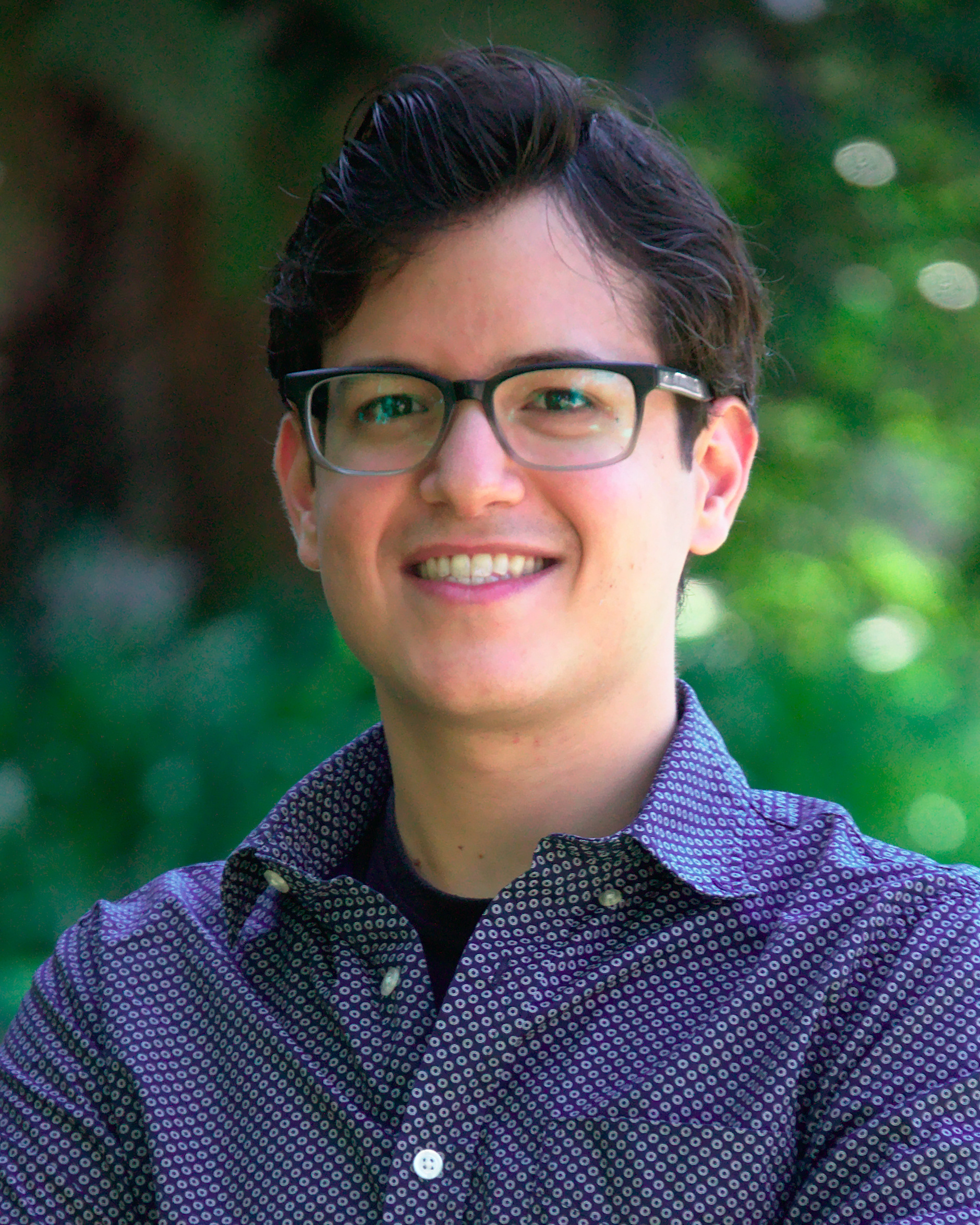}}]{Emmanuel Garza}
    received his B.S. in Engineering Physics from Tecnol\'{o}gico de Monterrey, Mexico, in 2013, and his PhD from the California Institute of Technology in 2020. He was named a Computing Innovation Fellow in 2020 by the Computing Research Association (CRA) and the Computing Community Consortium (CCC), and is currently a Postdoctoral Scholar at the Ming Hsieh Department of Electrical and Computer Engineering, University of Southern California. His research interests include boundary integral methods for electromagnetics, simulation and optimization of photonic devices, and high-performance computing.
\end{IEEEbiography}


\vspace{-4.4in}

\begin{IEEEbiography}[{\includegraphics[width=1in,height=1.25in,clip,keepaspectratio]{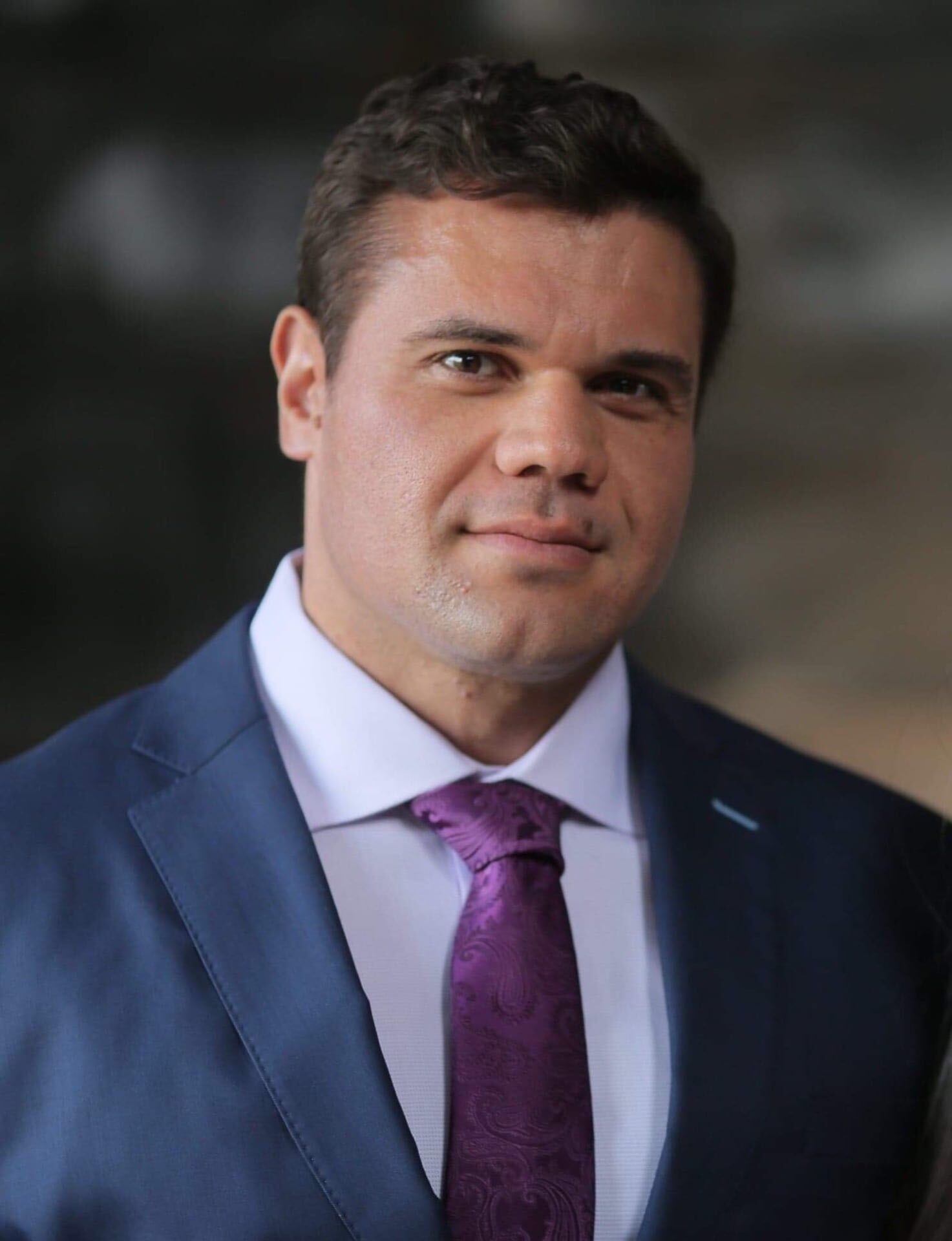}}]{Constantine Sideris}
    is an Assistant Professor of Electrical and Computer Engineering at the University of Southern California. He received the B.S., M.S., and PhD degrees with honors from the California Institute of Technology in 2010, 2011, and 2017 respectively. He was a visiting scholar at UC Berkeley’s Wireless Research Center from 2013 to 2014. He was a postdoctoral scholar in the Department of Computing and Mathematical Sciences at Caltech from 2017 to 2018 working on integral equation methods for electromagnetics. His research interests include RF and millimeter-wave integrated circuits for bioelectronics and wireless communications, applied electromagnetics, and computational electromagnetics for antenna design and nanophotonics. He was a recipient of the AFOSR YIP award in 2020, the Caltech Leadership Award in 2017, and an NSF graduate research fellowship in 2010.
\end{IEEEbiography}

\end{document}